\documentclass[longauth]{aa}
\pdfoutput=1
\usepackage[utf8]{inputenc}
\usepackage[OT2,T1]{fontenc}
\newcommand\textcyr[1]{{\fontencoding{OT2}\fontfamily{wncyr}\selectfont #1}}
\usepackage[english]{babel}
\usepackage{graphicx}
\usepackage{natbib}
\usepackage{verbatim}
\usepackage{textcomp}
\usepackage{color}
\usepackage{caption}
\usepackage{amsmath}
\DeclareCaptionFormat{cont}{#1 continued.#2#3\par}
\newcommand*\zsim{z$\sim$}

\newcommand*\micron{{\textmu{m}}}
\newcommand*\liravg{\langle L_{IR}\rangle}

\begin{document}

\title{The ALPINE-ALMA [CII] Survey: Obscured Star Formation Rate Density and Main Sequence of star-forming galaxies at z>4}
\author{Y.~Khusanova~(\textcyr{Ya. Khusanova})\inst{\ref{LAM},\ref{MPIA}} 
\and M.~Bethermin\inst{\ref{LAM}}
\and O.~Le~F\`evre\inst{\ref{LAM}} 
\and P.~Capak\inst{\ref{IPAC},\ref{DAWN}} 
\and A.~L.~Faisst \inst{\ref{IPAC}}
\and D.~Schaerer \inst{\ref{GenevaObs}}
\and J.~D.~Silverman\inst{\ref{KavliT},\ref{TokyoUni}}
\and P.~Cassata\inst{\ref{UniPadova}}
\and L.~Yan \inst{\ref{CaltechObs}}
\and M.~Ginolfi \inst{\ref{GenevaObs}} 
\and Y.~Fudamoto\inst{\ref{GenevaObs}}
\and F.~Loiacono\inst{\ref{BolognaUni},\ref{BolognaObs}}
\and R.~Amorin \inst{\ref{Serena1},\ref{Serena2}}
\and S.~Bardelli \inst{\ref{BolognaObs}}
\and M.~Boquien \inst{\ref{CITEVA}}
\and A.~Cimatti \inst{\ref{BolognaUni},\ref{FirenzeObs}}
\and M.~Dessauges-Zavadsky \inst{\ref{GenevaObs}}
\and C.~Gruppioni \inst{\ref{BolognaObs}}
\and N.~P.~Hathi \inst{\ref{STSI}}
\and G.~C.~Jones\inst{\ref{Cavendish},\ref{KavliC}}
\and A.~M.~Koekemoer \inst{\ref{STSI}}
\and G.~Lagache \inst{\ref{LAM}}
\and R.~Maiolino\inst{\ref{KavliC}}
\and B.~C.~Lemaux\inst{\ref{Davis}} 
\and P.~Oesch \inst{\ref{GenevaObs}}
\and F.~Pozzi \inst{\ref{BolognaUni}}
\and D.~A.~Riechers\inst{\ref{Cornell},\ref{MPIA}}
\and M.~Romano\inst{\ref{UniPadova},\ref{ObsPadova}}
\and M.~Talia\inst{\ref{BolognaObs},\ref{BolognaUni}}
\and S.~Toft \inst{\ref{DAWN},\ref{NBI}} 
\and D.~Vergani \inst{\ref{BolognaObs}}
\and G.~Zamorani \inst{\ref{BolognaObs}}
\and E.~Zucca \inst{\ref{BolognaObs}}
}
\institute{Aix Marseille Universit\'e, CNRS, LAM (Laboratoire d'Astrophysique de Marseille) UMR 7326, 13388, Marseille, France\label{LAM}
\and Max-Planck-Institut f\"{u}r Astronomie, K\"{o}nigstuhl 17, D-69117 Heidelberg, Germany\label{MPIA}
\and IPAC, California Institute of Technology, MC 314-6, 1200 E. California Blvd., Pasadena, CA, 91125, USA\label{IPAC}
\and Cosmic Dawn Center (DAWN)\label{DAWN}
\and Observatoire de Gen`eve, Université de Genève, 51 Ch. des Maillettes, 1290 Versoix, Switzerland \label{GenevaObs}
\and Kavli Institute for the Physics and Mathematics of the Universe, The University of Tokyo, Kashiwa, Japan 277-8583 \label{KavliT}
\and Department of Astronomy, School of Science, The University of Tokyo, 7-3-1 Hongo, Bunkyo, Tokyo 113-0033, Japan \label{TokyoUni}
\and Dipartimento di Fisica e Astronomia, Università di Padova, Vicolo dell’Osservatorio 3, I-35122, Padova, Italy \label{UniPadova}
\and Caltech Optical Observatories, California Institute of Technology, Pasadena, CA 91125, USA \label{CaltechObs}
\and Universit\`a di Bologna, Dipartimento di Fisica e Astronomia, Via Gobetti 93/2, I-40129, Bologna, Italy\label{BolognaUni}
\and Istituto Nazionale di Astrofisica: Osservatorio di Astrofisica e Scienza dello Spazio di Bologna, via Gobetti 93/3, 40129, Bologna, Italy\label{BolognaObs}
\and Instituto de Investigaci\'on Multidisciplinar en Ciencia y Tecnolog\'ia, Universidad de La Serena, Ra\'ul Bitr\'an 1305, La Serena, Chile\label{Serena1}
\and Departamento de Astronom\'ia, Universidad de La Serena, Av. Juan Cisternas 1200 Norte, La Serena, Chile\label{Serena2}
\and Centro de Astronomía (CITEVA), Universidad de Antofagasta, Avenida Angamos 601, Antofagasta, Chile \label{CITEVA}
\and INAF - Osservatorio Astrofisico di Arcetri, Largo E. Fermi 5, I-50125, Firenze, Italy \label{FirenzeObs}
\and Space Telescope Science Institute, 3700 San Martin Drive, Baltimore, MD 21218, USA \label{STSI}
\and Cavendish Laboratory, University of Cambridge, 19 J. J. Thomson Ave., Cambridge CB3 0HE, UK \label{Cavendish}
\and Kavli Institute for Cosmology, University of Cambridge, Madingley Road, Cambridge CB3 0HA, UK \label{KavliC}
\and Department of Physics and Astronomy, University of California, Davis, One Shields Ave., Davis, CA 95616, USA\label{Davis}
\and Department of Astronomy, Cornell University, Space Sciences Building, Ithaca, NY 14853, USA \label{Cornell}
\and INAF–Osservatorio Astronomico di Padova, Vicolo dell’Osservatorio 5, I-35122, Padova, Italy \label{ObsPadova}
\and Niels Bohr Institute, University of Copenhagen, Lyngbyvej 2, 2100 Copenhagen, Denmark \label{NBI}
}

\abstract{
Star formation rate (SFR) measurements at z>4 have relied mostly on rest-frame far-ultraviolet (FUV) observations. The corrections for dust attenuation based on IRX-$\beta$ relation are highly uncertain and are still debated in the literature. Hence, rest-frame far-infrared (FIR) observations are necessary to constrain the dust-obscured component of the SFR. In this paper, we exploit the rest-frame FIR continuum observations collected by the ALMA Large Program to INvestigate [CII] at Early times (ALPINE) to directly constrain the obscured SFR in galaxies at 4.4<z<5.9. We use stacks of continuum images to measure average infrared (IR) luminosities taking into account both detected and undetected sources. Based on these measurements, we measure the position of the main sequence of star-forming galaxies and the specific SFR (sSFR) at $z\sim4.5$ and $z\sim5.5$. We find that the main sequence and sSFR do not evolve significantly between $z\sim4.5$ and $z\sim5.5$, as opposed to lower redshifts. We develop a method to derive the obscured SFR density (SFRD) using the stellar masses or FUV-magnitudes as a proxy of FIR fluxes measured on the stacks and combining them with the galaxy stellar mass functions and FUV luminosity functions from the literature. We obtain consistent results independent of the chosen proxy. We find that the obscured fraction of SFRD is decreasing with increasing redshift but even at $z\sim5.5$ it constitutes around 61\% of the total SFRD.
}


\titlerunning{ALPINE: Infrared Star Formation Rate Density at z>4}
\authorrunning{Yana Khusanova et al.}

\maketitle

\section{Introduction}

\subsection{Star formation rate density at z>4}
\label{intro1}

The evolution of the star formation rate density (SFRD) is one of the fundamental elements for properly describing the history of the Universe. This measure provides information about the \emph{in situ} growth of galaxies, provides an inference for the number of ionizing photons produced by galaxies, and, hence, can help to place constraints on the history of the reionization \citep[e.g.][]{madau_cosmic_2014,bouwens_reionization_2015,Robertson2015}. 

The SFRD is generally measured by converting the dust-corrected FUV luminosity density of galaxies \citep[see e.g.][]{madau_cosmic_2014}. Since a fraction of the far-ultraviolet (FUV) light emitted by massive short-lived stars can be absorbed by dust, when present, the total SFRD consists of two components: the $SFRD_{FUV,uncorr}$ measured from the rest-frame FUV luminosity density, and the dust-hidden component $SFRD_{IR}$ measured from the rest-frame infrared (IR) luminosity density.

The ratio between these two components changes with time and, hence, redshift. The obscured fraction, $SFRD_{IR}$, dominates at low redshifts; it is 3 to 10 times higher than $SFRD_{FUV,uncorr}$ at z<3 \citep[e.g.,][]{Sanders2003, Takeuchi2003, Magnelli2011, Magnelli2013, Burgarella2013}. At higher redshift, the $SFRD_{IR}$ is poorly constrained. However, since metals are produced by stars, the first galaxies should have very low dust content, slightly enriched only by Population III stars. Therefore, in these galaxies, $SFRD_{FUV,uncorr}$ should be well in excess of the dust-hidden fraction $SFRD_{IR}$. As star formation continues, stars internal to the galaxy begin to generate more dust. While the survival of dust grains is regulated by a series of constructive and destructive processes \citep{Calura2017}, the net result is an increase of the  $SFRD_{IR}$ in respect to the $SFRD_{FUV,uncorr}$, but it is still unknown how fast the $SFRD_{IR}$ increases at early times and when it overcomes $SFRD_{FUV,uncorr}$, mainly due to the lack of large, statistically-representative galaxy populations observed and selected both in the FUV and far-infrared (FIR) at high redshift. 

Only a few attempts to constrain the $SFRD_{IR}$ at high redshifts exist in literature and their results remain inconclusive. \cite{Gruppioni2013} measured $SFRD_{IR}$ up to \zsim4 using Herschel selected galaxies and found that the $SFRD_{IR}$ decreases rapidly at z>3. \cite{Rowan-Robinson2016} used the 500 \textmu{}m sources from Herschel and found that the $SFRD_{IR}$ at z=3-6 is even higher than the estimates from the rest-frame UV, although these results still have significant uncertainty. The measurements of  \cite{Koprowski2017} using SCUBA-2 data at redshift up to z$\sim$5, on the other hand, suggest a very steep decrease of $SFRD_{IR}$, steeper than \cite{Gruppioni2013} \citep[see discussion in][]{Gruppioni2019}, so that $SFRD_{IR}$ at z>4 is even lower than $SFRD_{FUV,uncorr}$. This may be difficult to reconcile with \cite{Rowan-Robinson2016} estimates as well as with individual detections of very dusty galaxies at redshifts $z\simeq7$ \citep{Strandet2017, bowler_obscured_2018}, $z=8.312$ \citep{Tamura2019} and $z=8.38$ \citep{Laporte2017}, where the IR contribution should be extremely low if the $SFRD_{IR}$ continues to decrease rapidly with an increasing redshift. As the discoveries of individual highly dust-obscured objects are not from volume-complete samples, it is difficult to determine whether such objects are common in the early Universe. However, the recent study using a volume-complete sample from CO Luminosity Density at High Redshift (COLDz) survey by \cite{Riechers2020} resulted in detection of three massive dusty starburst galaxies over $\sim200,000$ Mpc$^3$, indicating a high space density of such dusty objects at z>5. Since this study was limited to luminous starburst galaxies, this is only a lower limit and less luminous galaxies can contribute even more to $SFRD_{IR}$. On the other hand, the previously mentioned studies based on measuring IR LFs \citep{Gruppioni2013,Rowan-Robinson2016, Koprowski2017} suffered from several deficiencies at the high-redshift end ($3.0 < z < 4.2$). At these redshifts, the observations were found to be considerably incomplete, and, furthermore, those galaxies that were observed were mostly brighter than typical galaxies at these redshifts. As such, conclusions related to the relative contribution of unobscured and obscured star formation at these epochs were limited in nature.

To overcome the difficulty of computing IR LFs at z>4, the measurements of the cosmic infrared background (CIB) produced by dusty galaxies can be used together with the modeling of dusty galaxy clustering, from which the $SFRD_{IR}$ can be then derived. \cite{Maniyar2018} did this analysis using the CIB anisotropy measurements and their cross-correlation with cosmic microwave background (CMB) lensing \citep[see also][]{Planck_CIB_2014}. The results turned out to be consistent with a shallower evolution of the $SFRD_{IR}$ at high redshifts compared to the results by \cite{Gruppioni2013} and  \cite{Koprowski2017}. At z>4, the $SFRD_{IR}$ does not drop below $SFRD_{FUV,uncorr}$ but approaches it up to \zsim6. 

Due to the difficulty in constraining the $SFRD_{IR}$ at redshifts z>4, the measurements of the total SFRD rely mostly on the rest-frame FUV part of the spectrum. To account for the obscured fraction, the relation between the infrared excess (IRX) and the UV spectral slope ($\beta$) \citep[][]{Meurer1999} is often used. However, stacking analysis \citep{Fudamoto2017}, as well as individual observations of high redshift galaxies with ALMA revealed significant scatter of dust properties. Most galaxies show a deficit of FIR flux compared to local galaxies with the same $\beta$ slopes \citep{Capak2015,Bouwens2016,Fudamoto2020}, which can be partially explained by higher dust temperatures \citep{Faisst2017}. The measurements of dust temperatures are scarce but indeed higher than in local galaxies (30-43 K at z$\sim$5) \citep{Faisst2020}. Since measurements of the SFRs at high redshifts are based on dust corrections coming from the IRX-$\beta$ relation, the scatter in the relation induces uncertainties on these measurements. For this reason, rest-frame FIR observations at z>4 remain important to constrain the IR luminosities independently.

\subsection{Main sequence and sSFR at z>4}

It is well established in the literature that star-forming galaxies form a so-called main sequence, a relationship between the amount of stellar mass in situ ($M_*$) and the SFR \citep{Noeske2007, Daddi2007, whitaker_star-formation_2012, speagle_highly_2014, Salmon2015, tasca_evolving_2015, schreiber_herschel_2015, Tomczak2016, santini_star_2017, pearson_main_2018, Khusanova2020VUDS}. However, the measurements of the SFRs of individual galaxies at z>4 suffer from the uncertainty of the IRX-$\beta$ relation and/or the dust temperature of the galaxies (see Sect. \ref{intro1}), hence, lower number of detection in the rest-frame FIR than expected assuming local IRX-$\beta$ \citep[e.g.,][]{Capak2015,Bouwens2016}. Besides, getting a wide volume in the rest-frame FIR at z>4 is expensive observationally. Therefore, the constraints on the main sequence of galaxies are limited and are mostly based on the rest-frame FUV data. 

Only a few attempts have been made to constrain the main sequence using sub-mm observations, mainly using data from Herschel \citep{schreiber_herschel_2015, pearson_main_2018}. At z>4, however, the measurements are limited by confusion noise \citep{Nguyen2010}, which makes it difficult to observe individual sources without careful and well-informed de-blending schemes. Therefore, these works primarily are limited to only the high mass end of the main sequence ($\log(M_*/M_{\odot})>10.5$). Star-forming galaxies with such high masses are rather rare, and it remains unclear if the slope of the main sequence remains the same over the whole range of masses.

The main sequence evolution with redshift is linked to the sSFR evolution defined as $SFR/M_*$, although the sSFR evolution can vary depending on the mass if the slope of the main sequence is not unity, as claimed in several studies \citep[e.g.,][]{Salmon2015, schreiber_herschel_2015, tasca_evolving_2015, Khusanova2020VUDS}. In that case, the measurements from UV selected samples, which probe galaxies with lower stellar masses ($\log(M_*/M_{\odot})\lesssim 10$) are not comparable with FIR samples probing higher stellar masses \citep{schreiber_herschel_2015}. The sSFR constraints at high redshifts are inconclusive: while many studies find a shallower evolution at high redshifts \citep{Bouwens_uv_2012, Gonzalez2014, tasca_evolving_2015, Faisst2016, santini_star_2017}, some find higher sSFR, consistent with steeper evolution, as at lower redshifts \citep{stark_keck_2012,deBarros2014}.

The ALMA Large Program to INvestigate [CII] at Early times (ALPINE) was designed with a goal of measuring the dust-hidden SFR in galaxies at z>4 from a representative UV selected sample \citep{Bethermin2020, lefevre_alpine_2019, Faisst2020}. In this paper, we use a large sample of 118 galaxies observed in the rest-frame FIR with ALMA to determine the $SFRD_{IR}$ at z>4 from ALPINE. We also determine their average and individual SFRs and constrain the main sequence over the stellar mass range $9\lesssim\log(M_*/M_{\odot})\lesssim11$. The combination of the rest-frame FUV and FIR data allows estimating SFRs without relying on IRX-$\beta$ relation and by directly measuring the total infrared luminosities. We present the ALPINE data in Sect. \ref{data_section}. In Sect. \ref{methods_section}, we describe the stacking code and the methods to measure the SFRD. In Sect \ref{sfrd_meas_sect}, we present and discuss our SFRD measurements. Sect. \ref{alp_ms_sect} is dedicated to a discussion about the main sequence and the sSFR evolution with redshift. In Sect. \ref{concl_section} we summarize our results.

Throughout the paper, we use a $\Lambda$CDM cosmology with $\Omega_{\Lambda}=0.70$, $\Omega_m=0.30$ and $h=0.7$. All magnitudes are given in the AB system. We assume a \cite{Chabrier2003} initial mass function (IMF) to convert luminosities to SFRs.

\section{Data}
\label{data_section}

ALPINE was designed to observe [CII] emission and the rest-frame FIR continuum of a representative sample of star-forming galaxies in the redshift range $4.4<z<5.$8 with a gap at redshifts 4.65<z<5.05, where the [CII] 158 \micron~emission line falls into the low transmission window \citep{lefevre_alpine_2019}. The selection of galaxies covers a wide range of stellar masses $9\lesssim\log(M_*/M_{\odot})\lesssim11$ and star formation rates derived from the rest-frame FUV with SED fitting $1\lesssim \log(SFR [M_{\odot}/yr])\lesssim3$ \citep[][]{Faisst2020}. The targets are ``normal'' star-forming galaxies and, based on observations in the rest-frame FUV, populate the main sequence. We use stacking of both detections and non-detections to derive the average properties of the sample (see Sect. \ref{methods_section}). In this way, we avoid bias towards Luminous Infrared Galaxies (LIRGs) and Ultra Luminous Infrared Galaxies (ULIRGs), compared to other observations of samples selected in the rest-frame FIR at these redshifts.

In total, 118 galaxies were observed in the period from May 2018 to January 2019 \citep[see][]{Bethermin2020}. The targets were primarily selected from two large spectroscopic surveys: VIMOS UltraDeep Survey \citep[VUDS,][]{lefevre_vimos_2015, Tasca2017} and DEep Imaging Multi-Object Spectrograph \citep[DEIMOS,][]{Faber2003} 10K Spectroscopic Survey \citep{DEIMOS_release_paper_2018}. All targets have secure spectroscopic redshifts from the rest-frame UV spectra and are imaged with both space and ground-based instruments \citep{Grogin2011, koekemoer_candels:_2011, Laigle2016, Skelton2014}. The secure spectroscopic redshift ensures that [CII], if present, will fall within the relatively narrow observed ALMA bandpass.

The ALMA data were calibrated with the standard ALMA pipeline. The CASA software was used for the data reduction \citep{McMullin2007}. To produce continuum maps, we excluded all the channels that correspond to the position of the [CII] line. In case of non-detections of the [CII] emission, we used spectroscopic redshifts from VUDS or DEIMOS 10K and excluded data within 500 km/s around the expected position of [CII] line. For galaxies with detected [CII] emission, we manually defined the window around the detected emission line. We then produced the continuum maps, using the task \texttt{tclean}. We used natural weighting to maximize SNR. More detail on data reduction can be found in \cite{Bethermin2020}.

The surveys was optimized to detect [CII] line emission. Therefore, the expected SNR for the rest-frame FIR continuum varies in the range from 0.5 to 3 for most targets based on the SED models of \cite{Bethermin2017}. We reached a sensitivity of 50 \textmu{}Jy/beam at 4.4<z<4.6 and 28\textmu{}Jy/beam at 5.1<z<5.8. We found that the 95\% purity is reached with a SNR=3.5 cut for targeted galaxies. Only 23 out of 118 (19\%) of the ALPINE targets are detected with such significance. However, given the known positions of the targets, we can extract the average continuum flux of non-detected galaxies using stacking, as described in Sect. \ref{stacking_section}.

The galaxies in ALPINE are chosen in the two well-studied fields, Cosmic Evolution Survey \citep[COSMOS, 89\% of targets,][]{Scoville2007_cosmos_overview} and Extended Chandra Deep Field-South Survey \citep[ECDFS, 11\% of targets,][]{Giacconi2002} that is covered by CANDELS \citep{koekemoer_candels:_2011, Grogin2011}. \cite{Faisst2020} analyzed the multi-wavelength ancillary data in both fields and derived the physical properties of galaxies using SED fitting (stellar masses, SFRs, rest-frame FUV luminosities, etc). The SED fitting is performed with the LePhare package \citep{Arnouts1999, ilbert_accurate_2006}, using \cite{bruzual_stellar_2003} models. The model parameters are outlined in Table \ref{sedfit_params}. The photometry from the public COSMOS catalog \citep{Laigle2016} was used for targets in the COSMOS field, and the 3D-HST catalog \citep{Skelton2014} was used for the ECDFS. The fitting was performed in flux density space and emission lines are included. More detail about the SED fitting of ALPINE galaxies can be found in \cite{Faisst2020}.

\begin{table}
	\caption{SED fitting model parameters}
	\label{sedfit_params}
	\centering
	\begin{tabular}{p{3cm} p{4.5cm}} 
		\hline\hline 
		\multicolumn{2}{c}{Input parameters} \\ \hline    
		IMF & \cite{Chabrier2003} \\
		SFH & Exponentially declining ($\tau=0.1,0.3,1.0,3.0$ Gyr) \\
		 & Delayed exponential ($\tau=0.1,0.5,1.0,3.0$ Gyr)\\
		 & Constant \\
		Attenuation law & Calzetti starburst \\
		E(B-V)$_s$ & 0, 0.05, 0.1, 0.15, 0.2, 0.25, 0.3, 0.35, 0.4, 0.45, 0.5 \\ 
		Metallicities & 0.2 $Z_{\odot}$ and 1.0 $Z_{\odot}$ \\ \hline    
	\end{tabular}
\end{table}

\section{Methods}
\label{methods_section}

\subsection{Stacking of the rest-frame FIR continuum images}
\label{stacking_section}

As mentioned above, for $\sim$81\% of the ALPINE targets, the observed continuum flux cannot be measured since it falls below the 3.5$\sigma$ limits. Therefore, we used stacking to measure the average flux of galaxies with similar physical properties. We use all the targeted galaxies including both detections and non-detections, since our goal is to get the average flux of the whole population of galaxies in each bin.

Here we use the mean stacks of continuum images. We only use the targeted galaxies and their UV-positions for alignment. Since the targets are not perfectly aligned on stacks due to uncertainties in astrometry or real physical offsets, the flux is measured using an aperture of 3 arcsec diameter. \cite{Bethermin2020} has tested three other methods of continuum flux measurements in addition to aperture photometry: Gaussian fitting on a continuum map and on the \textit{UV} plane, and peak flux. They all were proven to be consistent with each other, except for the peak flux. This is because the sources are not point-like. The other methods give equivalent results. However, on the stacks only aperture photometry can be used, since we cannot align the non-detected sources and their sizes are possible also different. Therefore, their mean profile will not be Gaussian. The aperture photometry, on the other hand, makes it possible to measure the flux without any assumption on the profile shape. Since the beam sizes in ALPINE do not exceed 1.6 arcsec, the 3 arcsec aperture is large enough to include all the flux from the targets if the sources are significantly extended. \cite{Fujimoto2020}, examined the radial surface brightness profiles of ALPINE targets and found that the sources are compact in the rest-frame FIR.

We use bootstrapping analysis to ensure the robustness of the measurements of the flux and to estimate the uncertainties coming from both the noise and the sample variance. Each stack is produced from a set of continuum images. To measure the average flux and its uncertainty, we randomly draw a new sample of images without withdrawal from the initial set and produce a new stack image, from which the continuum flux is remeasured. We repeat the procedure 1000 times and find the average and standard deviation of the measurement -- $\sigma_{boot}$. 

We also measure the flux in the field with the same aperture at each step. We mask the positions of the serendipitously detected objects and the center beforehand, to ensure that the flux measured in the field is not contaminated. The apertures are then placed at random positions, away from serendipitous sources. The variance of the field flux is an estimate of the photometric noise on the stacked image and is on average $\sigma_{phot}=0.05$ mJy for the typical number of stacked images (8). The average flux in the field is compatible with zero as expected ($F_{field}=0.1\pm50.0$ \textmu{}Jy).

As discussed in \cite{Bethermin2012}, the bootstrap uncertainty $\sigma_{boot}$ and the photometric noise $\sigma_{phot}$ are related as:

\begin{equation}
\label{sigma_boot_eq}
\sigma_{boot}=\frac{\sqrt{\sigma^2_{phot}+\sigma^2_{pop}}}{\sqrt{N_{stack}}},
\end{equation}
\noindent where $N_{stack}$ is the number of images in the stack and $\sigma_{pop}$ is the intrinsic population variance. The uncertainties derived using the bootstrap method are thus including both the photometric noise and the sample variance coming from the finite size and the possible large heterogeneity of the stacked samples.

\begin{figure}
	\resizebox{\hsize}{!}{\includegraphics{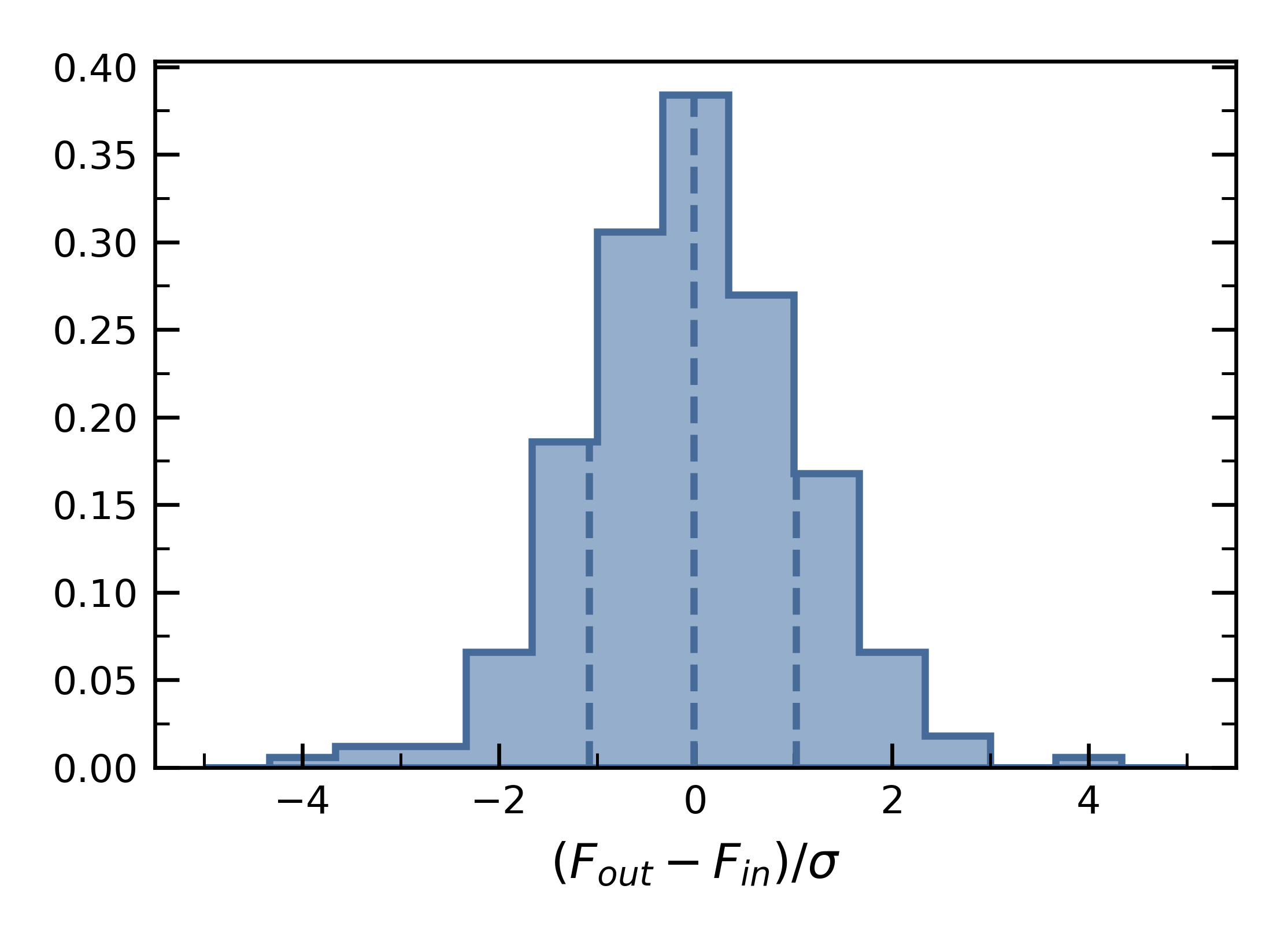}}
	\caption{Comparison of injected and recovered fluxes from stacking. The vertical dashed lines correspond to 16th, 50th and 84th percentiles of the distribution (-1.08, -0.02, 1.03). }
	\label{test_fig}
\end{figure}

In order to test reliability of the flux measurement on the stacks, we generated a number of fake sources on the images avoiding the center with fluxes drawn from a Gaussian distribution with a known mean $F_{in}$ and variance $\sigma_{in}$. The sources are Gaussian and their FWHM were allowed to vary from 0.8 to 1.6 arcsec \citep[as the beam sizes in ALPINE,][]{Bethermin2020}. We used the procedure described above to stack images with the fake sources, but in order to simulate position uncertainties of real targets, we added random offsets sampled from a Gaussian with $\sigma=0.12$ arcsec \citep[as the average astrometric offset of ALPINE targets,][]{Faisst2020}. Then we measured the average flux $F_{out}$ and the bootstrap uncertainty $\sigma_{out}$. We performed the procedure 250 times, varying the $F_{in}$ flux value in the range from 0.01 to 0.18 mJy or $\sim$0.25-3.5$\sigma_{phot}$ and the number of images from 6 to 13 (as we used later for the stacks).

Fig. \ref{test_fig} shows the distribution of $\Delta{F}/\sigma_{out}=(F_{out}-F_{in})/\sigma_{out}$. The mean of the distribution is -0.03 and the standard deviation is 1.13. The mean is compatible with zero. The standard deviation of the distribution can be reduced to $\sim1$, if we limit the analysis to stacks with a high number of images (>20), however, this is not feasible given the number of objects in the sample. We, therefore, make stacks with the number of objects in range from 6 to 13, but divide them in smaller bins to be able to constrain the $L_{IR}-M_*$ or $L_{IR}-M_{FUV}$ relation.

In conclusion, the average fluxes on stacks are well recovered within the error bars if the sources are not significantly extended or offset. This is the case for the detected targets \citep{Bethermin2020}, and stacking, therefore, can be safely used to recover the average continuum fluxes. 

\subsection{SFRD measurement: method description}
\label{methods_sfrd_section}

The total SFRD can be defined as \citep[see e.g.][]{madau_cosmic_2014}:
\begin{equation}
\label{sfrd_eq}
\begin{split}
SFRD=SFRD_{FUV,uncorr} + SFRD_{IR}=\\  \kappa_{FUV}\rho_{FUV}+\kappa_{IR}\rho_{IR},
\end{split}
\end{equation}
\noindent where $SFRD_{FUV,uncorr}$ is derived from the FUV luminosity density ($\rho_{FUV}$) not corrected for dust attenuation, and $SFRD_{IR}$ is derived from the IR luminosity density ($\rho_{IR}$). The conversion factors are $\kappa_{FUV}=1.5*10^{-10}M_{\odot}year^{-1}L_{\odot}^{-1}$ and $\kappa_{IR}=10^{-10}M_{\odot}year^{-1}L_{\odot}^{-1}$ for \cite{Chabrier2003} IMF \citep{kennicutt_global_1998}. 

\subsubsection{Measuring IR luminosity density}
\label{density_section}

\begin{figure*}
	\sidecaption
	\includegraphics[width=12cm]{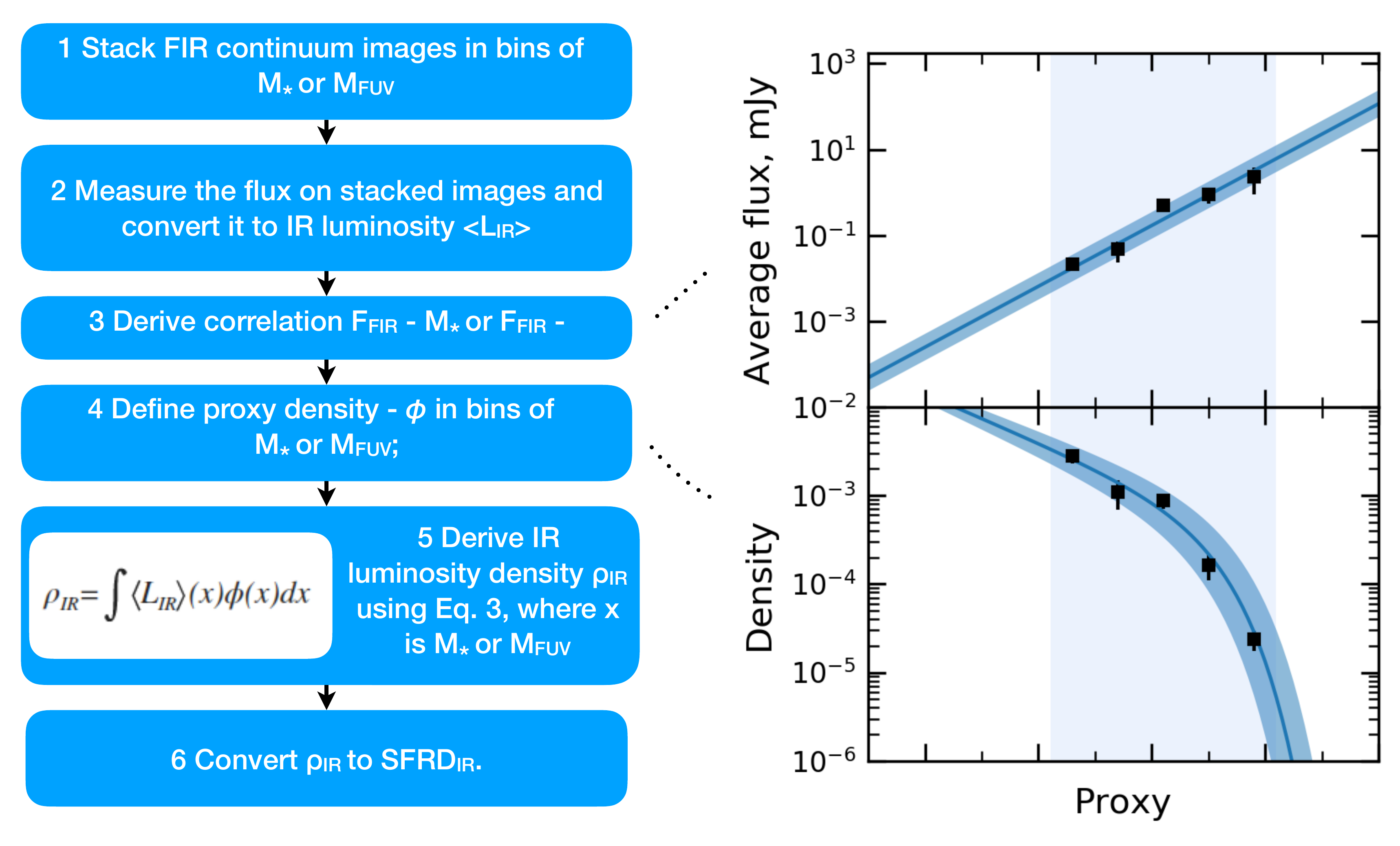}
	\caption{Illustration of the method used to determine the $SFRD_{IR}$. The left panel shows the steps performed. The right panel illustrates an arbitrary proxy density function (UVLF or GSMF, see text) and a power law relation between the property used as a proxy of FIR flux. The black squares represent measured values and the solid line the fit with uncertainties represented by a shaded region. The vertical shaded regions show the proxy range, in which the data is available for both the power law relation derived from ALPINE and the proxy density function.
}
	\label{method_illustation_fig_1}
\end{figure*}

Since ALPINE is not a volume-limited sample, it is rather difficult to determine the IR LF from the target sample \citep[see][for IR LF measurement based on non-target galaxies found in ALPINE]{Gruppioni2020}. We use, therefore, a different approach to obtain the IR luminosity density, which can then be converted to the $SFRD_{IR}$. The method is based on using some physical property as a proxy of the infrared luminosity - $L_{IR}$ and convolving the volume density distribution of this proxy with the mean LIR-proxy relation as:

\begin{equation}
\label{LDIR_formula}
\rho _{IR}=\int{ \liravg(x)\phi(x)dx},
\end{equation}
\noindent where $x$ is a proxy ($M_{FUV}$ or $M_{*}$, see Sect. \ref{proxy_choice_section}), $\liravg$ is the total infrared luminosity and $\phi(x)$ is either the FUV luminosity function (UVLF) or the galaxy stellar mass function (GSMF). The total IR luminosity is derived from the SED templates scaled to the 158 \textmu{}m flux measured on stacks and then integrated from 8 to 1000 \textmu{}m. We use average conversion factors $ \langle f_{temp, z\sim4.5} \rangle = 0.13 \pm 0.02$ and $\langle f_{temp, z\sim5.5} \rangle = 0.12 \pm 0.03$ derived from a set of templates from the literature, which are consistent with the stacked Herschel data (see Appendix \ref{sect:seds} for more detail on the templates and conversion factor). A summary of the method is presented in Fig. \ref{method_illustation_fig_1}.

The volume density distribution can be derived, for example, from the parent surveys, from which ALPINE targets were selected (VUDS or DEIMOS 10K) or from the literature measurements of the proxy density functions (UVLF or GSMF in our case). 

\subsubsection{The choice of proxy}
\label{proxy_choice_section}

The method described above relies on the assumption that there is a relation between the proxy and the IR luminosity. To first order, this is true for stellar masses ($M_{*}$, we will refer to "stellar mass" as "mass" hereafter) of star-forming galaxies, which sit on the main sequence. Their SFR increases with the mass \citep{whitaker_star-formation_2012,whitaker_constraining_2014,tasca_evolving_2015, Salmon2015, santini_star_2017,pearson_main_2018, Khusanova2020VUDS}. While it is still unclear what fraction of SFR is hidden by dust, we expect that there is a relation between the masses and the IR dust emission (see Sect. \ref{alp_ms_sect} for more discussion on the main sequence). Also the FUV-magnitudes are linked to SFR to first order. If we assume that more star-forming galaxies have higher IR luminosities, the FUV-magnitudes should correlate with the $L_{IR}$. Each proxy, however, has its own sources of uncertainty, which are discussed below. Therefore, we use both of them and compare the results.

\begin{figure*}
	\sidecaption
	\includegraphics[width=12cm]{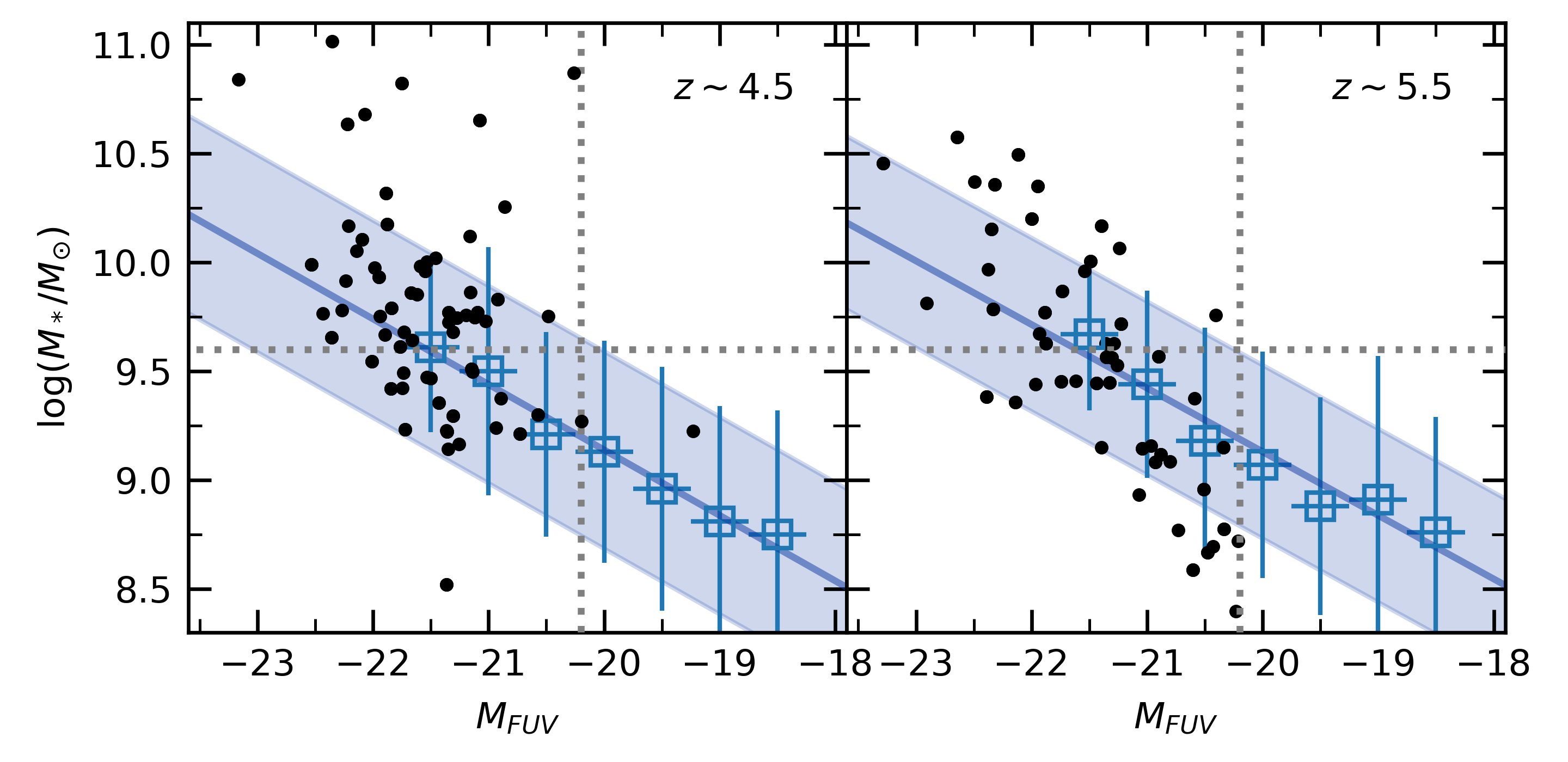}
	\caption{The $M_*-M_{FUV}$ relation. The black circles are ALPINE targets as measured in \citep{Faisst2020}. The open squares are measurements from \cite{Salmon2015}. The solid blue line is the fit to the measurements by \cite{Salmon2015} and the shaded areas are 1$\sigma$ uncertainties. The vertical dotted line shows the FUV luminosity cut used in selecting galaxies. The vertical dotted lines show the mass limits below which the flux measurements are considered to be upper limits. The left panel shows the \zsim4.5 redshift bin and the right panel shows the \zsim5.5 bin.}
	\label{mass_fuv}
\end{figure*}

The accuracy of the mass is limited, since it is derived from SED fitting and it is, therefore, model-dependent. 

The caveat in using the mass as a proxy is the fact that the sample of galaxies in ALPINE is selected to have FUV luminosities above $0.6L^*$. Hence, the galaxies selected at lower masses are not representative of typical galaxies at these masses, since they are biased towards higher UV luminosities at fixed mass. We show in Fig. \ref{mass_fuv} the median values of the $M_*-M_{FUV}$ relation from \cite{Salmon2015} and the fit of this relation. This relation agrees well with the $M_*$ and $M_{FUV}$ measurements for ALPINE targets. We find the mass at which the  1$\sigma$ envelope around the best fit crosses the FUV luminosity selection limit, where $\sigma$ is the scatter around the fit. As shown in Fig. \ref{mass_fuv}, the mass limit is $\log M_{lim}/M_{\odot}=9.6$ at both redshifts. Below this limit we are likely to miss galaxies with faint FUV magnitudes, biasing SFR towards higher values.

The FUV magnitudes are not affected by this bias. Since the spectroscopic redshifts are known, they are not model-dependent and can be derived with high accuracy given well-measured photometry \citep{Faisst2020}. We do not correct the derived FUV magnitudes for dust attenuation in order to keep them model-independent. However, the more dust is in the galaxy, the more FUV radiation is absorbed for a given SFR/IMF combination. Therefore, even if majority of faint in FUV galaxies have low dust content, very dust rich galaxies can contaminate FUV faint bins. In that case, the population will be heterogeneous in each bin and the relation may have larger scatter. This will make it more difficult to constrain the relation between FUV magnitude and IR luminosity. We determine the population variance with the bootstrapping procedure described in Sect. \ref{stacking_section} (see Eq. \ref{sigma_boot_eq}). The impact of the population variance is then propagated to the uncertainties on the SFRD. However, if the population variance is intrinsically too large, the use of FUV magnitudes as a proxy can be challenging to analyze small samples.

\subsubsection{Integration limits}

ALPINE probes only a range of masses ($9\lesssim\log(M_*/M_{\odot})\lesssim11$) or FUV-magnitudes ($-19\lesssim\log(M_*/M_{\odot})\lesssim-23$), used as a proxy (this range is schematically shown as shaded region in Fig. \ref{method_illustation_fig_1}). In this range, we have measurements of the rest-frame FIR flux directly from the stacks of ALPINE images. The proxy density is known from observations. The $SFRD_{IR}$ measured from integration in this range is, hence, the most robust lower limit of the dust-hidden fraction of SFRD. However, more infrared luminosity can be contained in FUV faint or low mass galaxies. While these galaxies are expected to have low dust content, they can still affect the results due to their high density in the Universe. Galaxies with high masses or bright in FUV are rare, but could also contribute to the total SFRD budget. Therefore, we need to extrapolate the observed in ALPINE $L_{IR}-proxy$ relation and the proxy density functions to obtain the estimate of the total $SFRD_{IR}$.

We determine, first, the lower limit ($SFRD_{IR,min}$) by integrating the Eq. \ref{LDIR_formula} within the observed in ALPINE mass of FUV-magnitude range. Then, for the extrapolation, we use the integration limits corresponding to $0.03L^*$ \footnote{This is a commonly used value in the literature \cite[e.g.][]{bouwens_uv_2015, madau_cosmic_2014, Khusanova2020VUDS}} and $100L^*$. These are $M_{FUV}=-17$ and $M_{FUV}=-26$, respectively. In order to define integration limits for masses, we convert these limits to SFRs. Then we use the relation between SFR and mass from the main sequence fit in \cite{Khusanova2020VUDS} to find the corresponding masses: $\log(M_*/M_{\odot})=6.0$ for the lower limit and $\log(M_*/M_{\odot})=12.4$ for the upper limit. With these limits, we can make a meaningful comparison between the $SFRD_{FUV,uncorr}$ derived in the literature from UV LFs and the $SFRD_{IR}$ measured here. 

We note that even after extrapolation, we can still be missing galaxies with high IR luminosity, but too faint in the rest-frame FUV to be detected in current photometric surveys \citep[e.g.,][]{Caputi2014, Franco2018, Williams2019, Wang2019, Gruppioni2020}. Their contribution will be discussed in the Sect. \ref{subsection_total_SFRD}.

\section{SFRD measurement results}
\label{sfrd_meas_sect}

\subsection{Proxy number density}
\label{proxy-density_section}

First, we need to define the proxy number density: UVLF or GSMF. Their measurements are widely available in the literature \citep[e.g., ][]{bouwens_uv_2015, Finkelstein2015, Ono2018, Pello2018, Khusanova2020VUDS, Song2016, Davidzon2017, Wright2018} but are still affected by many uncertainties. In particular, the faint and low mass end slopes are still not well constrained. Another uncertainty is the shape of the UVLF. While most authors assume the Schechter function form, some recent works found that at high redshift the double power law (DPL) is preferable form \citep{bowler_galaxy_2015, Ono2018, Khusanova2020VUDS}. We do not favor any particular UVLF or GSMF, but use several of them to show how the uncertainties on their parameters and functional form propagate into our SFRD estimates.

For the UVLF, we use the \cite{bouwens_uv_2015, Finkelstein2015, Ono2018, Pello2018} and \cite{Khusanova2020VUDS} results. The \cite{bouwens_uv_2015, Finkelstein2015} and \cite{Pello2018} UVLFs are based on large samples with photometric redshifts. The UVLF is represented with a Schechter function. \cite{Ono2018} uses a dropout selection partially confirmed by spectroscopic redshifts and \cite{Khusanova2020VUDS} uses the sample with spectroscopic redshifts. These studies probe the bright end of the UVLF and find a DPL form of UVLF.

We use \cite{Song2016, Davidzon2017} and \cite{Wright2018} measurements of GSMFs. \cite{Song2016} GSMFs are based on samples selected from CANDELS-GOODS field, \cite{Davidzon2017} used photometric selection of galaxies in COSMOS field, and \cite{Wright2018} used a joint analysis of GAMA, COSMOS, and 3D-HST data. We note that the physical parameters derived for ALPINE galaxies in COSMOS field were compared with the COSMOS15 catalogue and found to be consistent \citep[see ][]{Faisst2020}. Therefore, the uncertainties of mass measurements, which are more affected by the choice of the IMF and SFH in SED fitting compared to FUV magnitudes, should not have impact on our results, when we combine the COSMOS-based GSMF with LIR-mass relation from ALPINE.

\subsection{$\liravg$-proxy relation}
\label{lir-proxy_section}

 \begin{figure*}
	\centering
	\includegraphics[width=17cm]{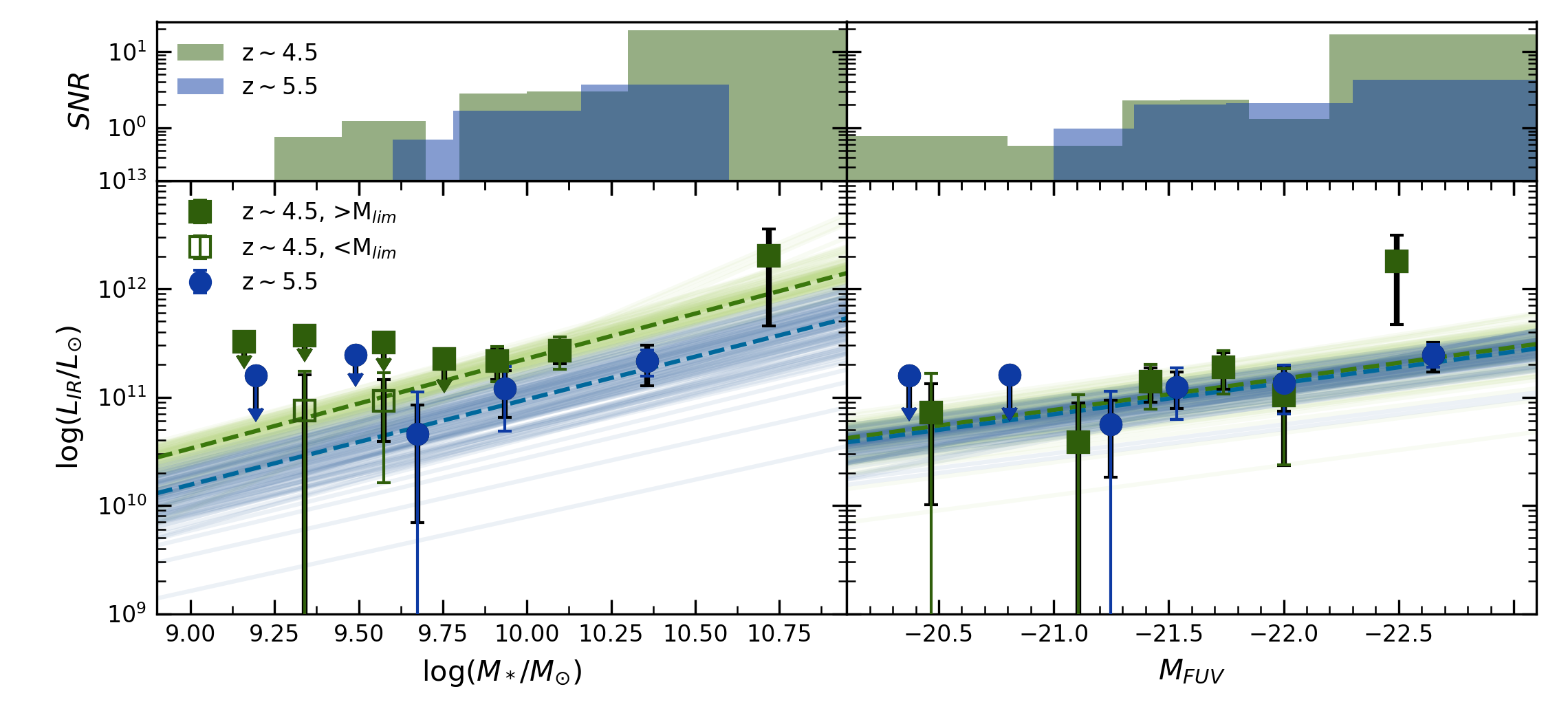}
	\caption{The $\liravg$-proxy relations: stellar mass as a proxy on the left panel and FUV magnitude on the right. The green squares and blue circles are the average IR luminosities on the stacks at $z\sim4.5$ and $z\sim5.5$, respectively. The black error bars show the population variance ($\sigma_{pop}$), the colored error bars show the photometric noise ($\sigma_{phot}$, see Eq. \ref{sigma_boot_eq}, Sect. \ref{stacking_section}). The open squares are the flux measurements on stacks below mass limit $M_{lim}$. The green and blue dashed lines are the best fits of the relation and the thin lines of corresponding color are the MCMC chain realizations of the fit. The top panels show the SNR on each stack.}
	\label{lir-proxy-fig}
\end{figure*}

As discussed in Sect. \ref{proxy_choice_section}, we use two proxies in this work: masses and the FUV magnitudes derived using the SED fitting as described in detail in \cite{Faisst2020}. The bin sizes are chosen to be small enough to reduce the heterogeneity of the underlying population but to include enough galaxies (6-13) in the bin to have robust average flux measurements. As was discussed in Sect. \ref{stacking_section}, even faint fluxes on the stacks give us information on the average flux of the underlying population of galaxies. Therefore, we use 3.5$\sigma_{phot}$ upper limits only when the measured fluxes of the stack have negative values. In all the remaining cases, we use the average flux measured with bootstrapping (the SNR in these cases is >0.5$\sigma_{phot}$, see Fig. \ref{lir-proxy-fig}). When stacking by mass bins, we consider the results as upper limits in bins below $\log M_{lim}/M_{\odot}=9.6$, for the reasons discussed in Sect. \ref{proxy_choice_section} (bias towards higher SFR because of the FUV selection of ALPINE). The upper limits are the sum of the flux measured on stacks and 3$\sigma$.

We converted the average rest-frame 160 \micron{} fluxes to $\liravg$ and found the $\liravg$-proxy relation. We show the resulting relations in Fig. \ref{lir-proxy-fig}. The $\liravg$ is clearly well correlated with both proxies. We fit this relation with a power law using the MCMC method with \texttt{emcee} package in python and checking at each step that the fit does not go above upper limits. The results are shown in Fig. \ref{lir-proxy-fig}. The slope is well constrained, and the normalization lies within the error bars. We use the MCMC chains later when deriving the $SFRD_{IR}$. This allows us to derive the uncertainty of the $SFRD_{IR}$ measurements, which relies on the precision of our $\liravg$-proxy relation.

In Fig. 4, we show the photometric errors and population variance computed as described in Sect. \ref{methods_section} (see Eq. \ref{sigma_boot_eq}) and SNR in each stack. The population variance at $z\sim4.5$ is $\sim0.91$ dex on average and at $z\sim5.5$ it is $\sim0.47$ dex, almost two times lower. As discussed in Sect. \ref{proxy_choice_section}, if the underlying population is heterogeneous, the scatter of its FIR properties will be large. This is what we observe at \zsim4.5. This makes the $L_{IR}-M_{FUV}$ relation particularly difficult to constrain at $z\sim4.5$. 

We note that the two points below $M_{lim}$, which were excluded from the fit due to a possible bias (see Sect. \ref{proxy_choice_section}), are in a very good agreement with the power law fit to the points above $M_{lim}$. Therefore, we trust these two points and use them later, when we derive the main sequence.

Interestingly, we observe a stronger evolution with redshift of the $L_{IR}-M_*$ relation than the $L_{IR}-M_{FUV}$ relation. As the redshift decreases, the $L_{IR}$ on average increases for galaxies with the same mass, which means the dust attenuation increases in these galaxies. 

\subsection{Infrared contribution to SFRD}

\begin{figure*}
	\includegraphics[width=12cm]{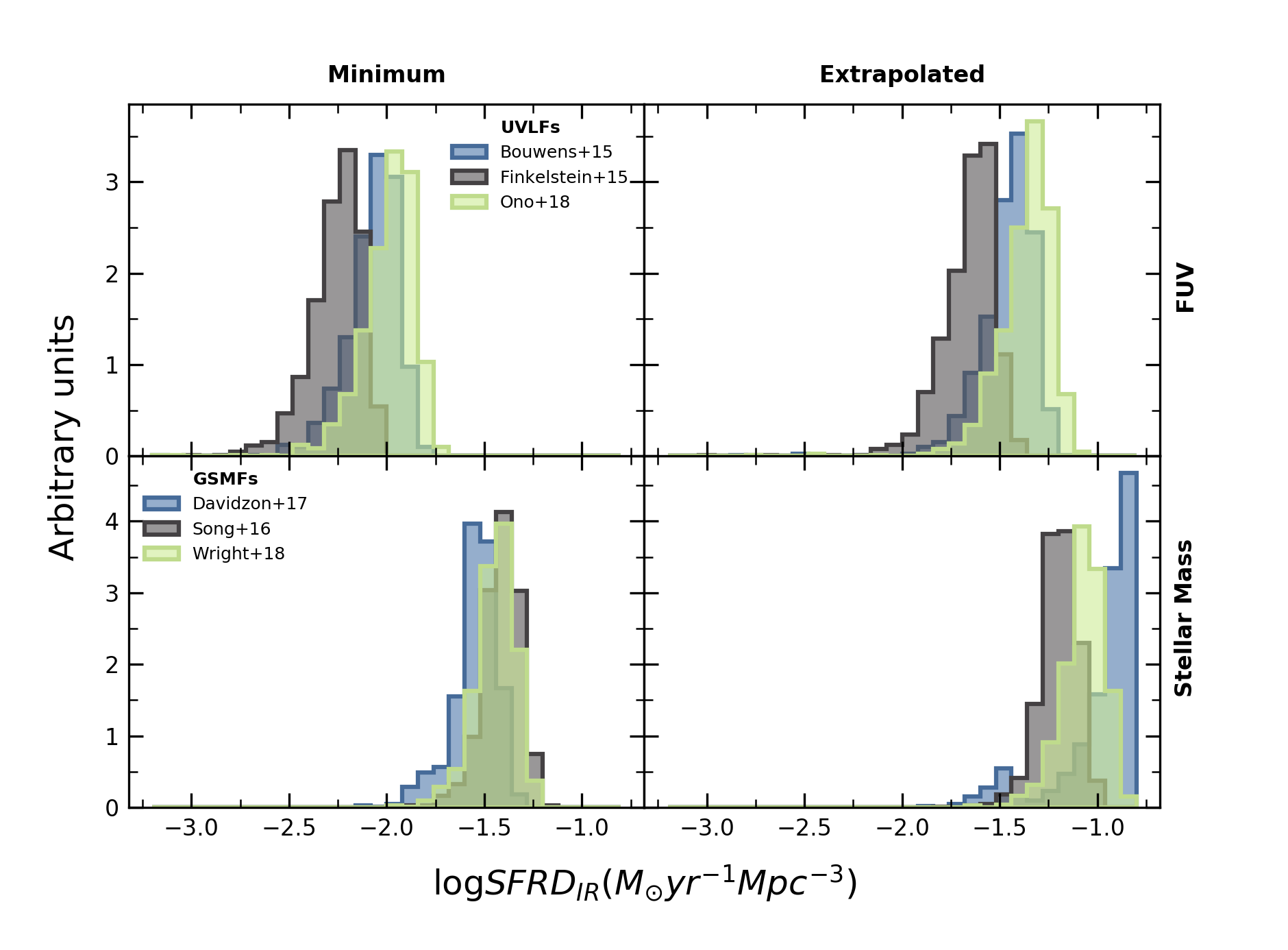}
	\caption{The PDF of the $SFRD_{IR}$ at redshift z$\sim$4.5. All PDF are derived with the same relations between the stellar mass (bottom panels) and FUV magnitudes (top panels). The minimum is derived in the observed by ALPINE range of stellar masses and FUV magnitudes and shown on the left panels. The SFRDs from extrapolated relation are shown on the right panels. Different colors correspond to different GSMFs and UV LFs \citep{bouwens_uv_2015, Finkelstein2015, Ono2018, Davidzon2017, Song2016,Wright2018}.}
	\label{hist_z4}
\end{figure*}

\begin{figure*}
	\includegraphics[width=12cm]{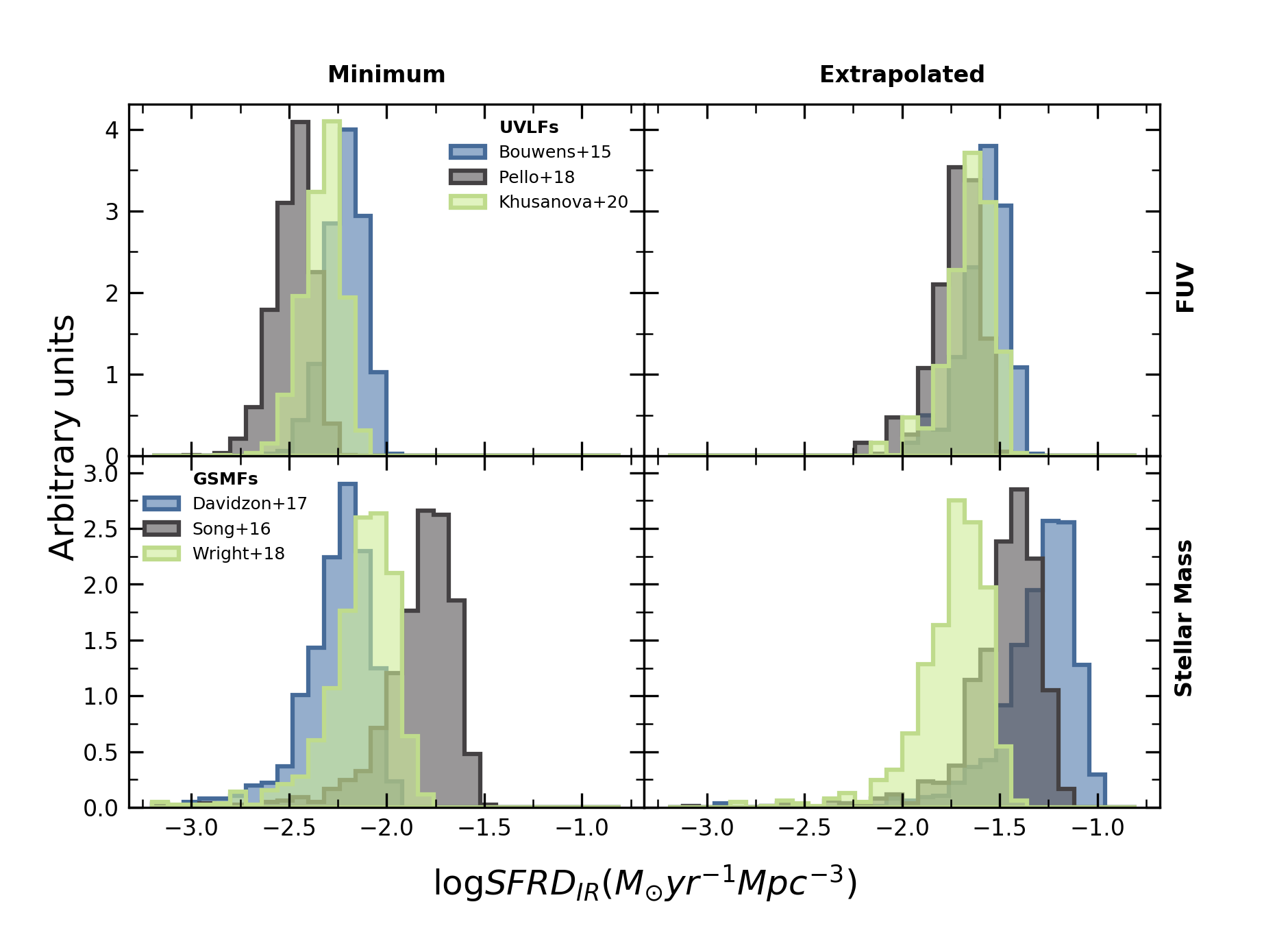}
	\caption{Same as Fig. \ref{hist_z4} but at redshift z$\sim$5.5. The UV LFs and GSMF are taken from \citep{bouwens_uv_2015, Pello2018, Khusanova2020VUDS, Davidzon2017, Song2016,Wright2018}.}
	\label{hist_z5}
\end{figure*}

We present the results of our measurements of $SFRD_{IR}$ in Fig. \ref{hist_z4} and \ref{hist_z5} for redshifts z$\sim$4.5 and z$\sim$5.5 respectively, and in Table \ref{sfrd_table}. Although the results depend on the choice of the proxy density parameters, in most cases they agree with each other within the error bars.

The choice of the proxy also plays a role in the constraining power of our method. At redshift \zsim4.5, the population variance in FUV bins is larger as we noted before. Hence, although we find a fit to the measured points, these constraints should be interpreted with caution. At \zsim5.5, on the contrary, we have weaker constraints when using the mass as a proxy. Due to the bias towards higher $FUV$ magnitudes in low mass bins, we had to exclude them from the fitting of the $L_{IR}-M_{*}$ relation. The remaining points have larger uncertainties at \zsim5.5. Hence, the fit of the $L_{IR}-M_{*}$ relation has larger uncertainties, which propagate into uncertainties of the SFRD estimates.

The measurements obtained with FUV as well as $M_*$ as a proxy agree well with each other within the errors in both redshift bins. The fact that the measurements obtained with different proxies are consistent with each other gives us on robustness of our constraints of the IR contribution to the SFRD. In the next section, we discuss our measurements in the context of the SFRD evolution and describe the results in more detail.

\begin{table}
\caption{Infrared SFRD ($\log{SFRD_{IR} M_{\odot}yr^{-1}Mpc^{-3}}$): Results obtained with various UVLFs and GSMFs from the literature. 
}
\label{sfrd_table}
\centering  
\begin{tabular}{c c c c}
	\hline\hline   
	\zsim4.5 & GSMF & Lower limit & Extrapolated \\
	\hline                          
	& Davidzon+17 & $-1.53_{-0.17}^{+0.16}$ & $-0.87_{-0.25}^{+0.18}$ \\
	& Song+16 & $-1.41_{-0.16}^{+0.16}$ & $-1.19_{-0.15}^{+0.17}$ \\
	& Wright+18 & $-1.43_{-0.17}^{+0.16}$  &  $-1.06_{-0.18}^{+0.17}$ \\
	& Combined & $-1.64$ & $-1.07_{-0.23}^{+0.28}$ \\
	\hline  
	& UVLF &  &  \\
	\hline
	& Bouwens+15 & $-2.05_{-0.20}^{+0.18}$ & $-1.44_{-0.21}^{+0.17}$ \\
	& Finkelstein+15 & $-2.24_{-0.20}^{+0.18}$ & $-1.64_{-0.21}^{+0.17}$ \\
	& Ono+18 & $-1.97_{-0.20}^{+0.18}$  &  $-1.34_{-0.21}^{+0.17}$ \\
	& Combined & $-2.44$ & $-1.48_{-0.25}^{+0.24}$ \\
	\hline
	\hline
	\zsim5.5 & GSMF &  &  \\
	\hline
	& Davidzon+17 & $-2.24_{-0.31}^{+0.26}$ & $-1.27_{-0.32}^{+0.26}$ \\
	& Song+16 & $-1.81_{-0.31}^{+0.26}$ & $-1.45_{-0.29}^{+0.26}$ \\
	& Wright+18 & $-2.11_{-0.31}^{+0.26}$ & $-1.72_{-0.30}^{+0.26}$ \\
	& Combined & $-2.54$ & $-1.50_{-0.39}^{+0.38}$ \\
	\hline
	& UVLF &  &  \\
	\hline
	& Bouwens+15 & $-2.20_{-0.20}^{+0.21}$ & $-1.56_{-0.23}^{+0.22}$ \\
	& Pello+18 & $-2.47_{-0.20}^{+0.21}$ & $-1.80_{-0.23}^{+0.22}$ \\
	& Khusanova+20 & $-2.32_{-0.20}^{+0.21}$ & $-1.73_{-0.23}^{+0.22}$ \\
	& Combined  & $-2.67$ & $-1.64_{-0.24}^{+0.23}$ \\
	\hline    
	\zsim4.5	& UVLF+GSMF & -2.44  & $-1.28_{-0.36}^{+0.38}$ \\
	\zsim5.5	& UVLF+GSMF & -2.67 & $-1.60_{-0.28}^{+0.36}$ \\
	\hline    
\end{tabular}
\end{table}

\subsubsection{The lower limit}
\label{lower_limit_section}

First, we obtained the lower limit on $SFRD_{IR}$. For that, we integrated the $\liravg$-proxy relation with the proxy density function in the range where both of them are known from the observations. Hence, this estimate is free from the uncertainties induced by extrapolations. The integration limits are $8.35\leq\log(M_*/M_{\odot})\leq10.5$ for masses and $-23.3\leq M_{FUV}\leq-20.0$ for FUV magnitudes.

As described in Sect. \ref{proxy-density_section} and \ref{lir-proxy_section}, each power law function describing $L_{IR}$-proxy relation is fitted with the MCMC (see Eq. \ref{LDIR_formula}). To derive the $SFRD_{IR}$ and its uncertainty, we use the MCMC chains of fit parameters to produce a resulting $SFRD_{IR}$ chain. In this way, we obtain a probability distribution function (PDF) of $SFRD_{IR}$, which is a normalized distribution of individual $SFRD_{IR}$ measurements at each step. These PDFs are shown in the left panel of Fig. \ref{hist_z4} and \ref{hist_z5} for both proxies. Depending on the choice of the proxy density function, the PDFs differ but remain consistent within the error bars for each proxy. In Fig. \ref{sfrd_uv_ir_fig}, we show the lowest value minus 1$\sigma$ as our lower limits for the IR contribution to SFRD. 

These are the most robust lower limits of the $SFRD_{IR}$ as they are obtained with a large and representative sample of galaxies at $4.4<z<5.8$ and only using the proxy range, where actual observations are available. Given this lower limit, the infrared contribution to the total SFRD is already at least 8-13\% at $z>4$ as compared to the total SFRD from the \cite{madau_cosmic_2014} curve. While this is seemingly a small amount, these results are in severe disagreement with the results by \cite{Koprowski2017}. They fitted the evolution of $SFRD_{IR}$ with redshift by the Gaussian function. The extrapolation of this curve gives $\log (SFRD_{IR} M_{\odot}yr^{-1}Mpc^{-3}) =-3.95$ at \zsim5.5, which is much lower than the ALPINE lower limit at this redshift $\log (SFRD_{IR} M_{\odot}yr^{-1}Mpc^{-3}) =-2.67$. Although, the agreement between ALPINE lower limits and $SFRD_{IR}$ measurements by \cite{Koprowski2017} could be reached by a shallower function at z>4, we note that we only considered a small range of masses or FUV magnitudes for computing the $SFRD_{IR}$ and our sample is UV-selected. Hence, the real values at both redshifts are even higher and in even stronger tension with \cite{Koprowski2017}. Due to the fact that the work of \cite{Koprowski2017} is based on the infrared luminosity functions and in the high redshift bins, the observations barely cover the faint end slope of the IR LF, whose uncertainty could be one of the source of this discrepancy.

\subsubsection{Extrapolated $SFRD_{IR}$}

\begin{figure*}
	\sidecaption
	\includegraphics[width=12cm]{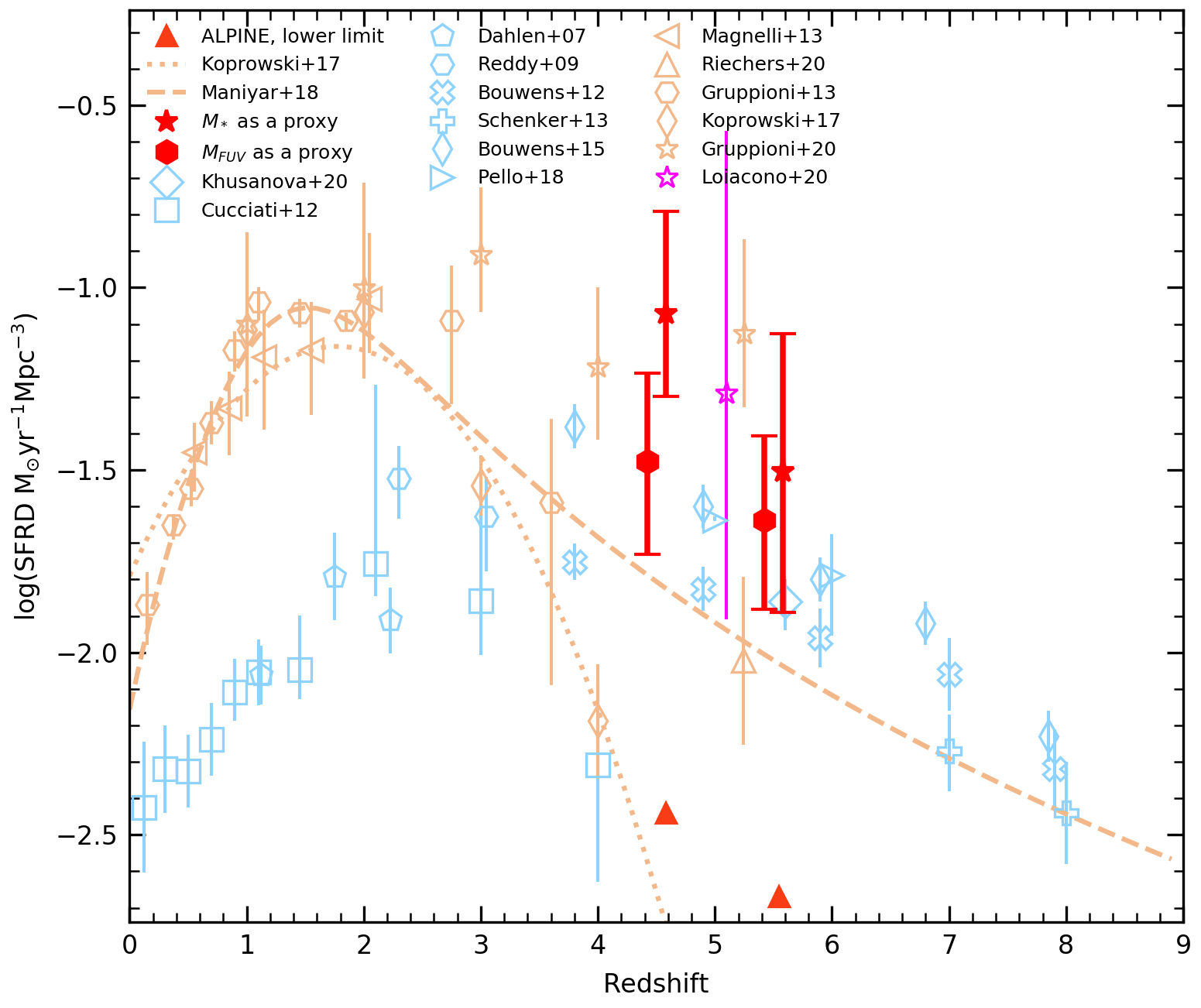}
	\caption{The $SFRD_{IR}$ and $SFRD_{FUV,uncorr}$ evolution with redshift. The red symbols are the $SFRD_{IR}$, derived in this paper. Small offsets of $\pm$0.08 are applied to their redshift for clarity. The triangles are the lower limits for $SFRD_{IR}$ derived in Sect. \ref{lower_limit_section}. The blue symbols are $SFRD_{FUV,uncorr}$ derived from the FUV data \citep{Cucciati2012, Dahlen2007, Reddy2009, Bouwens_uv_2012, Schenker2013, bouwens_uv_2015, Pello2018, Khusanova2020VUDS}, orange symbols are $SFRD_{IR}$ from  \cite{Magnelli2013,Gruppioni2013,Koprowski2017} and \cite{Gruppioni2020}. The magenta point is from [CII] luminosity function based on the field sample of \cite{Loiacono2020}. Orange lines are the fits of $SFRD_{IR}$ evolution from \cite{Koprowski2017} and \cite{Maniyar2018}.}
	\label{uv_uncor_ir_fig}
\end{figure*}

The total $SFRD_{IR}$ is higher than the lower limit derived above. To derive the total $SFRD_{IR}$, we assume that the $\liravg-proxy$ relation is the same at all FUV-magnitudes and masses and extrapolate the $\liravg-proxy$ relation as well as the proxy number density function to the integration limits, corresponding to $0.03L_*$ and $100L_*$ or $\log(M_*/M_{\odot})=6.0$ and $\log(M_*/M_{\odot})=12.4$ for masses, as mentioned in Sect. \ref{methods_sfrd_section}. We obtain the PDFs in the same way as above and show the results on the right panels in Fig. \ref{hist_z4} and \ref{hist_z5}. To place the results in the context of the SFRD evolution, we combine the PDFs obtained with different proxy density functions from the literature. Since the lengths are the same for all chains, combining them does not favor any particular proxy density function. We obtain an average estimate of the SFRD and our uncertainties naturally include the systematics coming from the GSMF and UVLF.

Fig. \ref{uv_uncor_ir_fig} illustrates the evolution  with redshift of $SFRD_{IR}$ and the uncorrected for dust estimate from FUV luminosity density - $SFRD_{FUV,uncorr}$. We observe a gradual decrease of the $SFRD_{IR}$, which only starts to intersect with the $SFRD_{FUV,uncorr}$ at z>5. Even at such high redshifts, the $SFRD_{FUV,uncorr}$ does not overcome the $SFRD_{IR}$. It remains an open question, however, at which redshift the $SFRD_{FUV,uncorr}$ overcomes the $SFRD_{IR}$, since at z>4 the $SFRD_{IR}$ is only gradually approaching the $SFRD_{FUV,uncorr}$. If the evolution is flatter at higher redshifts, then the dust contribution becomes negligible compared to $SFRD_{FUV,uncorr}$ only at very high redshifts. This trend is consistent with the extrapolation of the \cite{Maniyar2018} model of $SFRD_{IR}$ evolution, which is based on CIB anisotropies. Our results agree well with this model at both redshifts. However, if the contribution from ultra-dusty objects, which are missing from the optical and near-infrared (NIR) catalogs is significant, the $SFRD_{IR}$ evolution can be even flatter.

In this paper, we only use the target sources from ALPINE. Recently, \cite{Gruppioni2020} determined the $SFRD_{IR}$ using the non-target ALPINE sources. Our results are in good agreement with the results by \cite{Gruppioni2020} being higher. This is not surprising, since these sources were not selected from the surveys probing rest-frame UV. Therefore, they are better suited for tracing the the star formation contribution coming from ultra-dusty galaxies.

\subsubsection{The total SFRD}
\label{subsection_total_SFRD}

 \begin{figure*}
	\sidecaption
	\includegraphics[width=12cm]{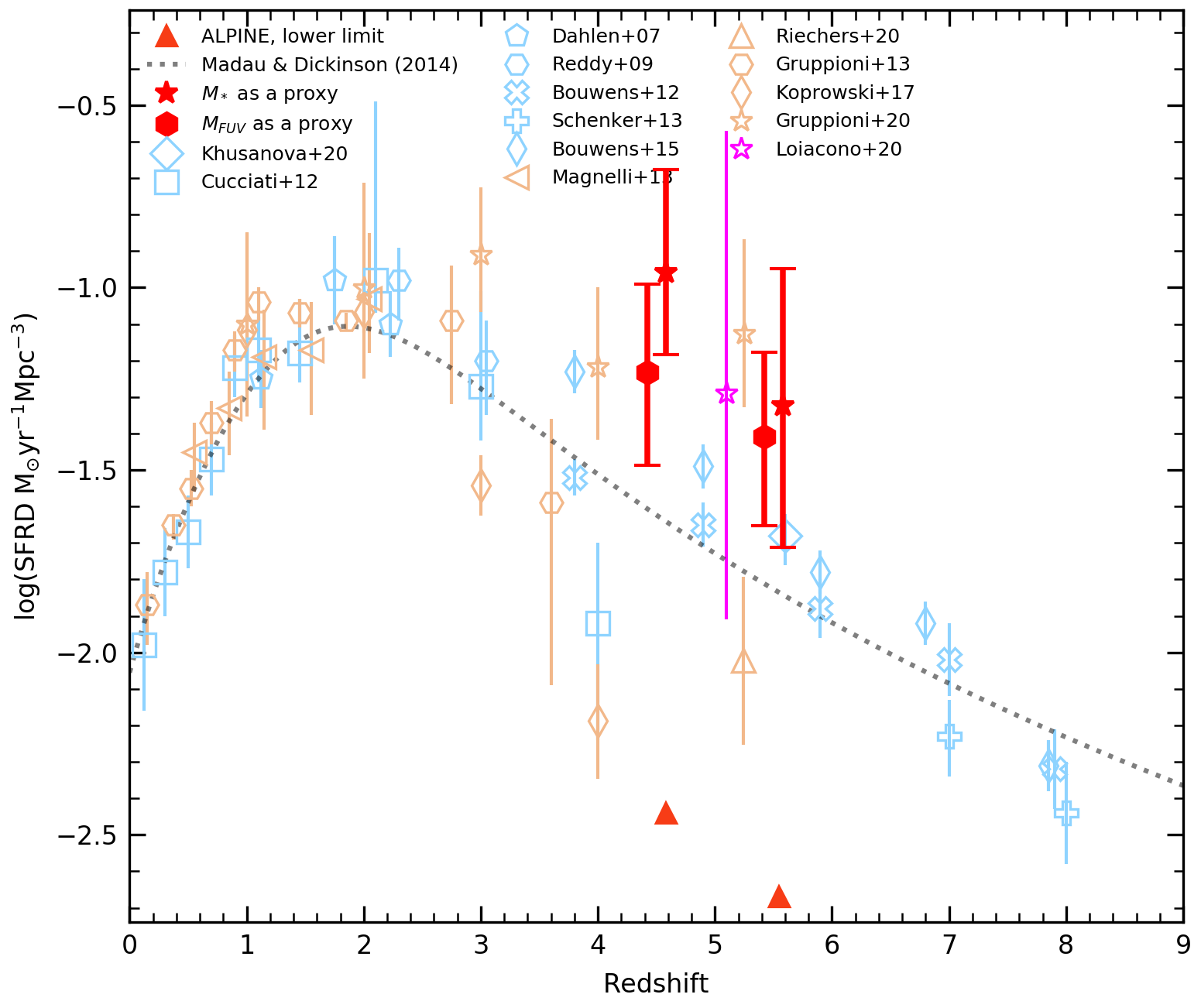}
	\caption{The evolution of the total SFRD (Eq. \ref{sfrd_eq}) with redshift. The red symbols are the total SFRDs, derived in this paper. Small offsets of $\pm$0.08 are applied to their redshift for clarity. The triangles are the lower limits for $SFRD_{IR}$ derived in Sect. \ref{lower_limit_section}. The blue symbols are dust-corrected SFRD derived from the FUV data \citep{Cucciati2012, Dahlen2007, Reddy2009, Bouwens_uv_2012, Schenker2013, bouwens_uv_2015, Pello2018, Khusanova2020VUDS}, orange symbols are derived from reradiated dust emission from forming stars as measured from the IR \citep{Magnelli2013,Gruppioni2013, Koprowski2017}. The magenta point is from [CII] luminosity function based on the field sample of \cite{Loiacono2020}. The dotted line is the fit of SFRD evolution from \cite{madau_cosmic_2014}. }
	\label{sfrd_uv_ir_fig}
\end{figure*}

We now define the total SFRD at \zsim4.5 and \zsim5.5 by summing the $SFRD_{UV,uncorr}$ with the $SFRD_{IR}$ from ALPINE. We plot our results in Fig. \ref{sfrd_uv_ir_fig}. The results are in agreement with the FUV based estimates corrected for dust attenuation from the literature within the error bars at both redshifts. They are also in agreement with the measurements based on non-target sources from ALPINE \citep{Gruppioni2020} and with the measurement based on the [CII] luminosity function of non-target sources of the field sample \citep{Loiacono2020}. However, our results obtained by a combination of two proxies (see Table \ref{sfrd_table}) are 0.5 dex and 0.3 dex higher than the fit of SFRD evolution by \cite{madau_cosmic_2014} or in 1.4$\sigma$ and 1.2$\sigma$ tension at \zsim4.5 and \zsim5.5 respectively. Although, given the uncertainty on the $SFRD_{IR}$ measurement, it is not possible to reach firm conclusions, the evolution of the total SFRD at z>4 is possibly shallower than the best fit in \cite{madau_cosmic_2014}.

\begin{figure}
	\resizebox{\hsize}{!}{\includegraphics{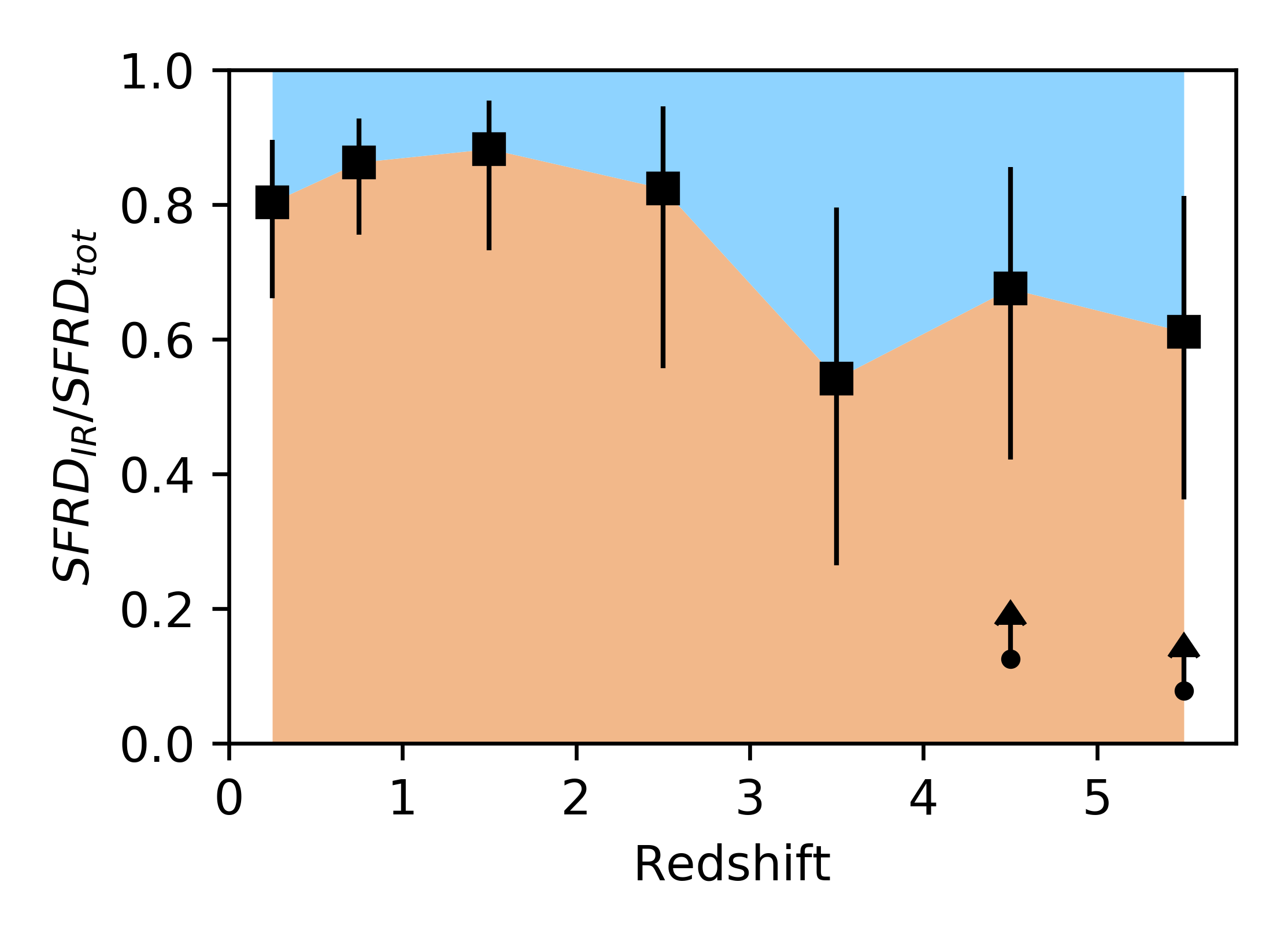}}
	\caption{The evolution of the fraction of the dust obscured $SFRD_{IR}$ as a function of redshift. The arrows show the lower limits obtained using the lower limits discussed in Sect. \ref{lower_limit_section}. The points at z<4 are obtained using a compilation of the literature from Fig. \ref{sfrd_uv_ir_fig}. The points at z>4 are obtained using the UVLF+GSMF combined measurements from Table \ref{sfrd_table}. }
	\label{fig_fraction}
\end{figure}

In Fig. \ref{fig_fraction}, we show the evolution of the dust obscured fraction of SFRD with redshift. We use all the points from the literature together with our measurements shown in Fig. \ref{uv_uncor_ir_fig} and \ref{sfrd_uv_ir_fig} to calculate the average $SFRD_{IR}$ and $SFRD_{UV,uncorr}$ in each redshift bin and the ratio of $SFRD_{IR}$ to the total SFRD. We find the infrared contribution to the total SFRD equal to $68_{-25}^{+18}\%$ and $61_{-25}^{+20}\%$ at \zsim4.5 and \zsim5.5 respectively. Although significant uncertainties remain, our results show that the dust obscured fraction of SFRD is significant even at z>4. Our results are consistent with semi-analytical model (SAM) by \cite{Cousin2019}, which predict that $\sim70\%$ of UV radiation is absorbed by dust at z=5, and with the best fit model by \cite{Zavala2018}, which predicts 35\%-85\% at z$\sim$4-5. 

Since we used a sample of galaxies detected in FUV, we did not take into account the contribution from highly obscured and purely IR galaxies, which remains unknown at these redshifts, mainly due to the difficulties in detecting such galaxies. The low number counts of the most massive galaxies makes it difficult to make a statistical census. No sources with z>5 were found by the deep ALMA observations of the Hubble Ultra-Deep Field \citep[HUDF;][]{Aravena2016, Dunlop2017, Gonzalez-Lopez2019}. Only recently, a few sources with no optical or NIR counterparts were discovered with ALMA at \zsim4 and \zsim5 \citep{Franco2018,Williams2019} and even \zsim6.9 \citep{Strandet2017, Marrone2018}. The largest sample up to date was assembled by \citet{Wang2019}. It contains 39 Hband dropouts at z>3 observed with ALMA. Given their results, such galaxies could have an additional contribution of $\sim11\%$ to the total SFRD at $z>4$. Since their sample is restricted to massive H-band dropouts, this is only a lower limit. The lower mass H-band dropouts could contribute to it significantly. Indeed, given the most recent results by \cite{Riechers2020}, the dust obscured SFRD is $\log (SFRD_{IR} M_{\odot}yr^{-1}Mpc^{-3}) =-2.02$ or  22\% of the total SFRD (based on the space density of two sources detected by CO (J=2$\rightarrow$1) and not detected in the rest-frame UV and optical).

Another uncertainty comes from the fact that the conversion from the 160 \micron{} flux to $L_{IR}$ in this work is based only on one point in the infrared SED. Therefore, our measurement of the total IR luminosity are dependent on the choice of the dust SED template. The infrared SED in high redshift galaxies is uncertain due to poorly constrained dust temperatures \citep{Pavesi2016, Bouwens2016, Faisst2017, Faisst2020}. We tested a number of templates against the stacking of Herschel data and only used the ones, which are compatible with the Herschel stacks at redshift range of ALPINE (see Appendix). The best fit of modified black body to these stacks gives the dust temperature $T_{dust}=41\pm1$ K at $z\sim4.5$ and $T_{dust}=43\pm5$ K at $z\sim5.5$ consistent with most recent measurements by \cite{Faisst2020} ($T_{dust}=38\pm8$ at $z\sim5.5$). Although some uncertainties on the dust temperature and the rest-frame FIR part of the SED remain, we incorporate them into the uncertainty of our SFRD measurement by using an average conversion factor between the templates compatible with the stacks of Heschel data and taking into account the uncertainty of the conversion factor. Nevertheless, more observations are needed to better constrain the dust temperatures and SED of galaxies at z>4.

\section{Star-forming main sequence and sSFR evolution}
\label{alp_ms_sect}

\begin{figure*}
	\sidecaption
	\includegraphics[width=12cm]{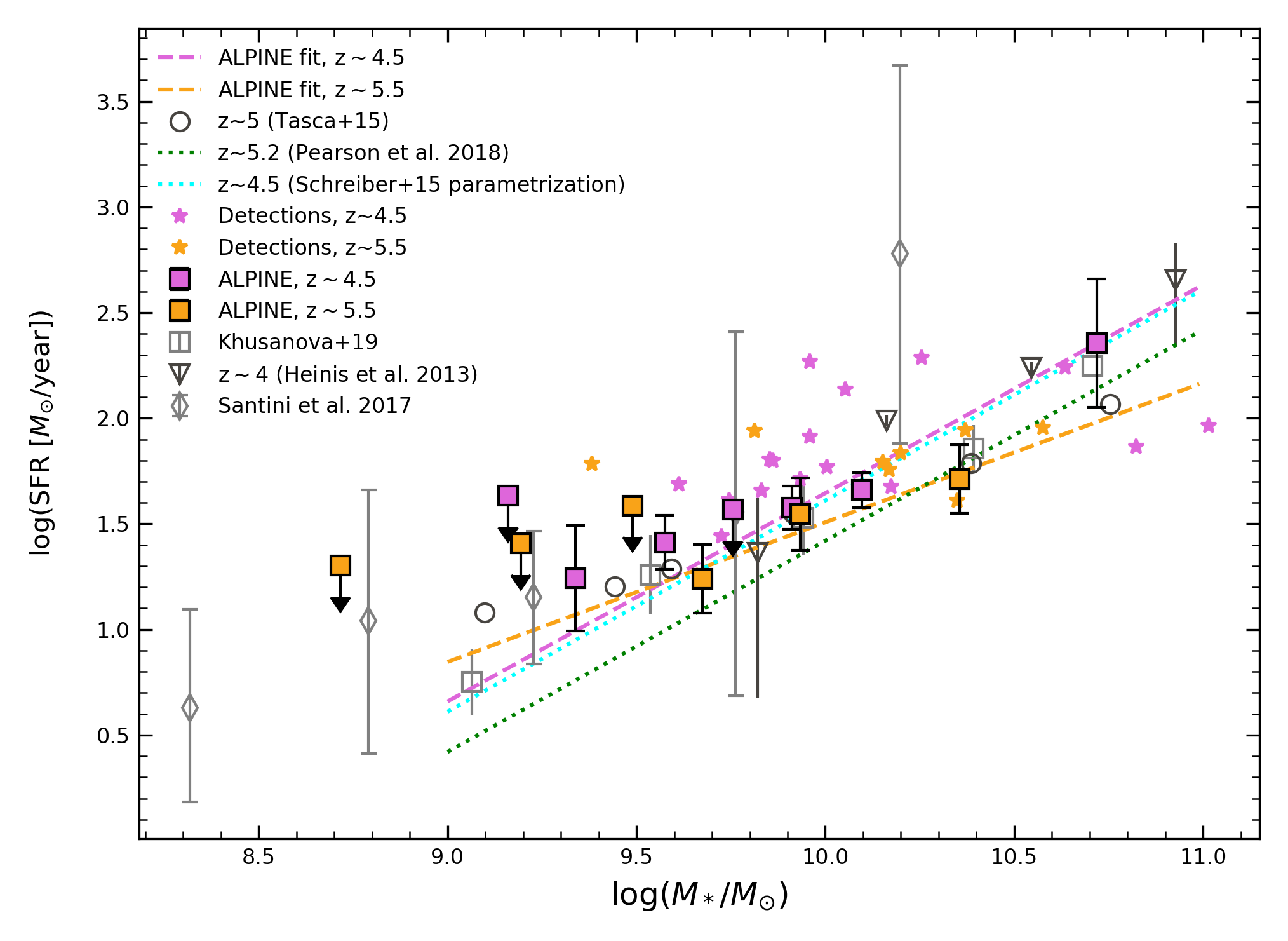}
	\caption{The main sequence of star-forming galaxies at \zsim4.5 and \zsim5.5. The filled squares are the measurements of average SFR from stacks; the open symbols are results from the literature; the magenta and orange lines are the fits of the main sequence at \zsim4.5 and \zsim5.5 respectively; the cyan and green lines are the fits of the main sequence from the literature. The colored stars are the individual galaxies in ALPINE with FIR continuum detections at >3$\sigma$.}
	\label{ms_alpine_fig}
\end{figure*}

\begin{figure*}
	\sidecaption
	\includegraphics[width=12cm]{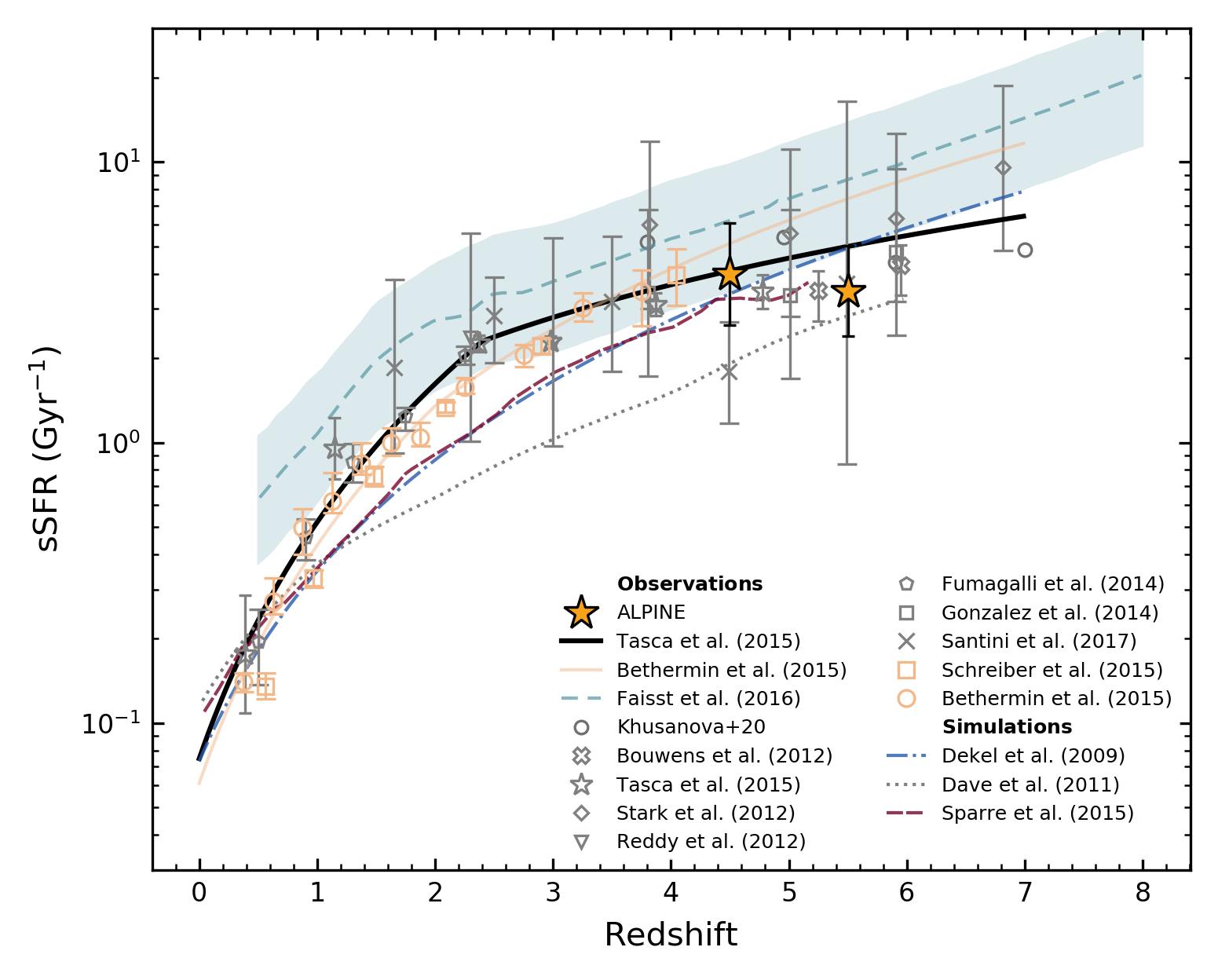}
	\caption{SSFR evolution with redshift. Results of various observations and numerical simulations are shown (gray symbols are based on FUV data and orange symbols are based on the rest-frame FIR data). The orange stars show this paper's results. The blue stars are the results obtained with VUDS data in \cite{Khusanova2020VUDS}. The solid line shows the fit to the previous results from VUDS \citep{tasca_evolving_2015}, the shaded area and the dashed line show the results coming from the observations of EW(H$\alpha$) in COSMOS \citep{Faisst2016}.}
	\label{ssfr_evol_fig}
\end{figure*}

In Sect. \ref{lir-proxy_section}, we derived the $\liravg-M_*$ relation. Here we use this relation to derive the main sequence, i.e. the relation between the mass and SFR for these galaxies. In each mass bin, we derive average SFR as $\langle SFR\rangle=\kappa_{FUV}\langle L_{1600}\rangle+\kappa_{IR}\langle L_{IR}\rangle$. The average $\langle L_{1600}\rangle$ is computed from FUV absolute magnitudes, determined with the SED fitting, and the average $\liravg$ from stacks. As discussed in Sect. \ref{proxy_choice_section}, we consider the results from the stacks as upper limits at masses below the completeness limits.

We derived the main sequence in the two redshifts bins: 4.4<z<4.7 and 5.1<z<5.8. We show the results in Fig. \ref{ms_alpine_fig}. The main sequence  is well-defined and is in very good agreement with previous results from the literature \citep{santini_star_2017,tasca_evolving_2015, Heinis2013, Khusanova2020VUDS}. We fitted the relation between SFR and mass with a power law and found a slope $\alpha=0.99\pm0.08$ at \zsim4.5. We find an excellent agreement with the \cite{schreiber_herschel_2015} parametrization of the main sequence evolution at \zsim4.5 based on Herschel data, even if we measure the SFRs at lower masses compared to \cite{schreiber_herschel_2015}.

At \zsim5.5, we obtain a shallower slope $\alpha=0.66\pm0.21$. The difference with the slope at \zsim4.5 is only $1.5\sigma$, providing no evidence for an evolution of the slope between the two redshift bins. We note that at \zsim5.5, we could only provide FIR flux upper limits at lower masses due to the selection effects (see Sect. \ref{proxy_choice_section}). Therefore, we only used three points to fit our data and the fit is subject to larger uncertainties. The slope of the main sequence found with Herschel data \citep{pearson_main_2018} at the same redshift but at larger masses is steeper than found with ALPINE data, but still consistent with our data on stacks within the error bars.

In Fig. \ref{ms_alpine_fig}, we also plot the individual measurements for galaxies with continuum detections. As expected, most of them are above the main sequence since they represent the brightest galaxies with high SFRs. This confirms that it is necessary to use stacking to obtain unbiased measurements of the average SFR as a function of $M_*$ and fit the main sequence.

We measured the sSFR in the mass range $9.6<\log(M_*/M_{\odot})<9.8$ as in \cite{santini_star_2017}. The average sSFR is $\log{sSFR}(Gyr^{-1})=-8.4\pm0.18$ and $\log{sSFR}(Gyr^{-1})=-8.45\pm0.16$ at \zsim4.5 and \zsim5.5 respectively. We present a comparison with results from the literature in Fig. \ref{ssfr_evol_fig}. We find good agreement within the error bars with the results from \cite{Khusanova2020VUDS} at z>5 and with Herschel data at \zsim4 \citep{Bethermin2015, schreiber_herschel_2015}, although we note that these works only probe the high mass end ($\log(M_*/M_{\odot})>10.0$) at \zsim4. The results are clearly against a steep increase of sSFR at z>4 and even indicate a possible decrease in sSFR at high redshift. More observations at higher redshifts are necessary to explore this trend.

If the growth of galaxies is regulated mainly by gas accretion through cold streams, the sSFR evolves as $\sim(1+z)^{2.25}$ \citep{dekel_cold_2009, dave_galaxy_2011, sparre_star_2015}. This is not the case at high redshifts. Therefore, another mechanism may play a role in the growth of galaxies, such as major and minor mergers \citep[see, e.g.,][]{Faisst2016}. Another possible explanation could be the suppression of star formation by stronger stellar feedback at high redshifts and reduced star formation efficiency \citep[see, e.g.,][]{Weinmann2011}. Future studies will place constraints on the role of different mechanisms regulating star formation and the growth of galaxies at high redshifts.

\section{Conclusions}
\label{concl_section}

In this paper, we explored the SFRD, sSFR and main sequence evolution at z>4 with ALPINE. Using the stacking technique, we constrained the average properties of galaxies even with low number of individual detections. 

We determined the robust lower limits $\log SFRD_{IR} (M_{\odot}yr^{-1}Mpc^{-3}) =-2.44$ and $\log SFRD_{IR} (M_{\odot}yr^{-1}Mpc^{-3}) =-2.67$ at \zsim4.5 and 5.5 respectively, which indicate that the dust plays an important role at high redshift  and that dust build-up happens quite rapidly during and just after epoch of reionization.

We then extrapolated the density functions and found the total dust-hidden IR contribution to SFRD using either masses or FUV magnitudes as a proxy of infrared luminosity. Using the masses, we obtain $\log SFRD_{IR,tot} (M_{\odot}yr^{-1}Mpc^{-3}) = -1.07^{+0.28}_{-0.23}$ and $\log SFRD_{IR,tot} (M_{\odot}yr^{-1}Mpc^{-3}) = -1.48^{+0.24}_{-0.25}$ at \zsim4.5 and 5.5 respectively. Using the FUV magnitudes, we obtain $\log SFRD_{IR,tot} (M_{\odot}yr^{-1}Mpc^{-3}) = -1.50^{+0.38}_{-0.39}$ and $\log SFRD_{IR,tot} (M_{\odot}yr^{-1}Mpc^{-3}) = -1.64^{+0.23}_{-0.24}$ at \zsim4.5 and 5.5 respectively. The results agree with each other within the error bars and suggest that $68_{-25}^{+18}\%$ and $61_{-25}^{+20}\%$ of the star formation is hidden by dust at z$\sim$4.5 and \zsim5.5 respectively. While at z<4 the $SFRD_{IR}$ component dominates over the $SFRD_{FUV,uncorr}$, at z>4 their contribution is comparable. The $SFRD_{IR}$ approaches and possibly crosses the $SFRD_{FUV,uncorr}$ at z>5 but this is yet to be confirmed with future observations.

The evolution of the total SFRD at redshifts higher than the peak of SFRD at \zsim2.5 may be shallower than previously measured. Further observations leading to reduced uncertainties are needed to confirm this conclusion.

We also provided robust measurements of SFRs of galaxies at z>4 and used them to determine the main sequence at \zsim4.5 and \zsim5.5. We found that the observed main sequence is in good agreement with the previous results based on the rest-frame UV and optical data, as well as the IR data for the most massive galaxies. The main sequence has a slope $\alpha=0.99\pm0.08$ at \zsim4.5 and $\alpha=0.66\pm0.21$ at \zsim5.5, and no signs of turn-over at high masses. No significant evolution is observed from \zsim4.5 to \zsim5.5.

We used the SFR and mass measurements to determine the average sSFR of galaxies at z>4. The results support a shallower or non-existent sSFR evolution at high redshifts than predicted from models of cold gas accretion, or no evolution. Other mechanisms should play an important role in governing the growth of galaxies.

\begin{acknowledgements} This paper is dedicated to the memory of Olivier Le~F\`evre, PI of the ALPINE survey. This work is based on data products made available at the CESAM data center, Laboratoire d'Astrophysique de Marseille, France. YK acknowledges the support by funding from the European Research Council Advanced Grant ERC--2010--AdG--268107--EARLY. AC, CG, FL, FP and MT acknowledge the support from grant PRIN MIUR 2017 - 20173ML3WW\_001. GL acknowledges support from the European Research Council (ERC) under the European Union’s Horizon 2020 research and innovation programme (project CONCERTO, grant agreement No 788212) and from the Excellence Initiative of Aix-Marseille University-A*Midex, a French “Investissements d’Avenir” programme. D.R. acknowledges support from the National Science Foundation under grant numbers AST-1614213 and AST-1910107. D.R. also acknowledges support from the Alexander von Humboldt Foundation through a Humboldt Research Fellowship for Experienced Researchers. G.C.J. acknowledges ERC Advanced Grant 695671 ``QUENCH'' and support by the Science and Technology Facilities Council (STFC). ST acknowledges support from the ERC Consolidator Grant funding scheme (project ConTExt, grant number No. 648179). The Cosmic Dawn Center is funded by the Danish National Research Foundation under grant No. 140. RA acknowledges support from FONDECYT Regular Grant 1202007.
\end{acknowledgements}

\bibliographystyle{aa}
\bibliography{references}

\begin{appendix}

\section{Conversion of FIR flux to IR luminosity}

\label{sect:seds}

To convert the rest-frame 160 \textmu{}m flux to IR luminosity, we scale our flux measurement to the dust SED template and then integrate it from 8 to 1000 \textmu{}m. Since we use stacks of galaxies, which are not on the same redshift, we use the mean redshift of galaxies in stacks for that conversion. Since we only obtain a measurement of one point in SED, our conversion to IR luminosity will depend on dust SED templates.

There are numerous dust SED templates available in the literature \citep[e.g.,][]{Magdis2012, Bethermin2017, AlvarezMarquez2016, Casey2018, Schreiber2018, DeRossi2018, AlvarezMarquez2019}. Some of them were already discussed and tested by \cite{Bethermin2020}. \cite{Bethermin2020} made stacks of Herschel data for galaxies at z>4 selected via photometric redshifts. They then compared the SED templates to these stacks. We define the conversion factor $f_{temp}$ for each template and the $\chi^2$ from comparison to \textit{Herschel} stacks at both redshifts (see Table \ref{table_templates}). The conversion factor is defined as $f_{temp} = \nu L_{\nu = 158 \mu m} / L_{IR}$. We discard the templates with reduced $\chi^2$ > 1.5 in at least one of the redshift bins (see Fig. \ref{fig_templates_discarded}). We define the mean conversion factors from the remaining templates $ \langle f_{temp, z\sim4.5} \rangle = 0.13 \pm 0.02$ and $\langle f_{temp, z\sim5.5} \rangle = 0.12 \pm 0.03$ for redshift bins 4<z<5 and 5<z<6 respectively. These are the values we use to convert the monochromatic flux to the infrared luminosity and propagate their uncertainties into our SFRD measurements.

We note that although the \cite{DeRossi2018} template passes our $\chi^2$ criteria, it goes above the upper limits in redshift bin 4<z<5, but is perfectly consistent at 5<z<6. Therefore, we use this template only for  5<z<6 redshift bin. 

\begin{table*}
	\caption{The dust SED templates, conversion factors $f_{temp}$ and $SFRD_{IR}$ corresponding to these templates.
	}
	\label{table_templates}
	\centering  
	\begin{tabular}{c c c c c c}
		\hline\hline   
		Reference & $f_{temp}$ & $SFRD_{IR, z=4.5}$ & $\chi^2_{z=4.5}$ & $SFRD_{IR, z=5.5}$ & $\chi^2_{z=5.5}$ \\
		\hline  
\cite{Bethermin2017}  & 0.133 & $-1.37_{-0.27}^{+0.21}$ &  1.61  & $-1.71_{-0.17}^{+0.21}$ &  3.40\\
\cite{AlvarezMarquez2019} L1  & 0.105 & $-1.47_{-0.27}^{+0.21}$ &  1.16  & $-1.81_{-0.17}^{+0.21}$ &  3.19\\
\cite{AlvarezMarquez2019} L2  & 0.152 & $-1.31_{-0.27}^{+0.21}$ &  4.69  & $-1.65_{-0.17}^{+0.21}$ &  5.36\\
\cite{AlvarezMarquez2019} L3  & 0.123 & $-1.40_{-0.27}^{+0.21}$ &  1.89  & $-1.74_{-0.17}^{+0.21}$ &  3.81\\
\cite{AlvarezMarquez2019} M2  & 0.119 & $-1.42_{-0.27}^{+0.21}$ &  1.04  & $-1.76_{-0.17}^{+0.21}$ &  3.35\\
\cite{AlvarezMarquez2019} M3  & 0.148 & $-1.32_{-0.27}^{+0.21}$ &  1.97  & $-1.66_{-0.17}^{+0.21}$ &  4.28\\
\cite{AlvarezMarquez2019} M4  & 0.152 & $-1.31_{-0.27}^{+0.21}$ &  2.20  & $-1.65_{-0.17}^{+0.21}$ &  4.46\\
\cite{AlvarezMarquez2019} M5  & 0.153 & $-1.31_{-0.27}^{+0.21}$ &  2.28  & $-1.65_{-0.17}^{+0.21}$ &  4.52\\
\cite{Schreiber2018}, z=4.5  & 0.093 & $-1.53_{-0.27}^{+0.21}$ &  2.12  & $-1.86_{-0.17}^{+0.21}$ &  - \\
\cite{Schreiber2018}, z=5.5  & 0.073 & $-1.63_{-0.27}^{+0.21}$ &  -  & $-1.97_{-0.17}^{+0.21}$ &  3.98\\
\cite{DeRossi2018}  & 0.067 & $-1.67_{-0.27}^{+0.21}$ &  4.65  & $-2.01_{-0.17}^{+0.21}$ &  3.76\\
\hline
Discarded templates & & & & & \\
\hline
\cite{AlvarezMarquez2016}     & 0.189 & $-1.22_{-0.27}^{+0.21}$ &  9.74  & $-1.56_{-0.17}^{+0.21}$ &  7.05\\
\cite{AlvarezMarquez2019} L4  & 0.229 & $-1.13_{-0.27}^{+0.21}$ &  12.89  & $-1.47_{-0.17}^{+0.21}$ &  8.54\\
\cite{AlvarezMarquez2019} M1  & 0.354 & $-0.94_{-0.27}^{+0.21}$ &  35.84  & $-1.28_{-0.17}^{+0.21}$ &  16.14\\
\cite{Casey2018} A  & 0.244 & $-1.11_{-0.27}^{+0.21}$ &  11.71  & $-1.45_{-0.17}^{+0.21}$ &  9.00\\
\cite{Casey2018} B  & 0.036 & $-1.94_{-0.27}^{+0.21}$ &  12.99  & $-2.23_{-0.17}^{+0.21}$ &  7.74\\
		\hline    
	\end{tabular}
\end{table*}

\begin{figure*}[h]
\centering
\includegraphics[width=14cm]{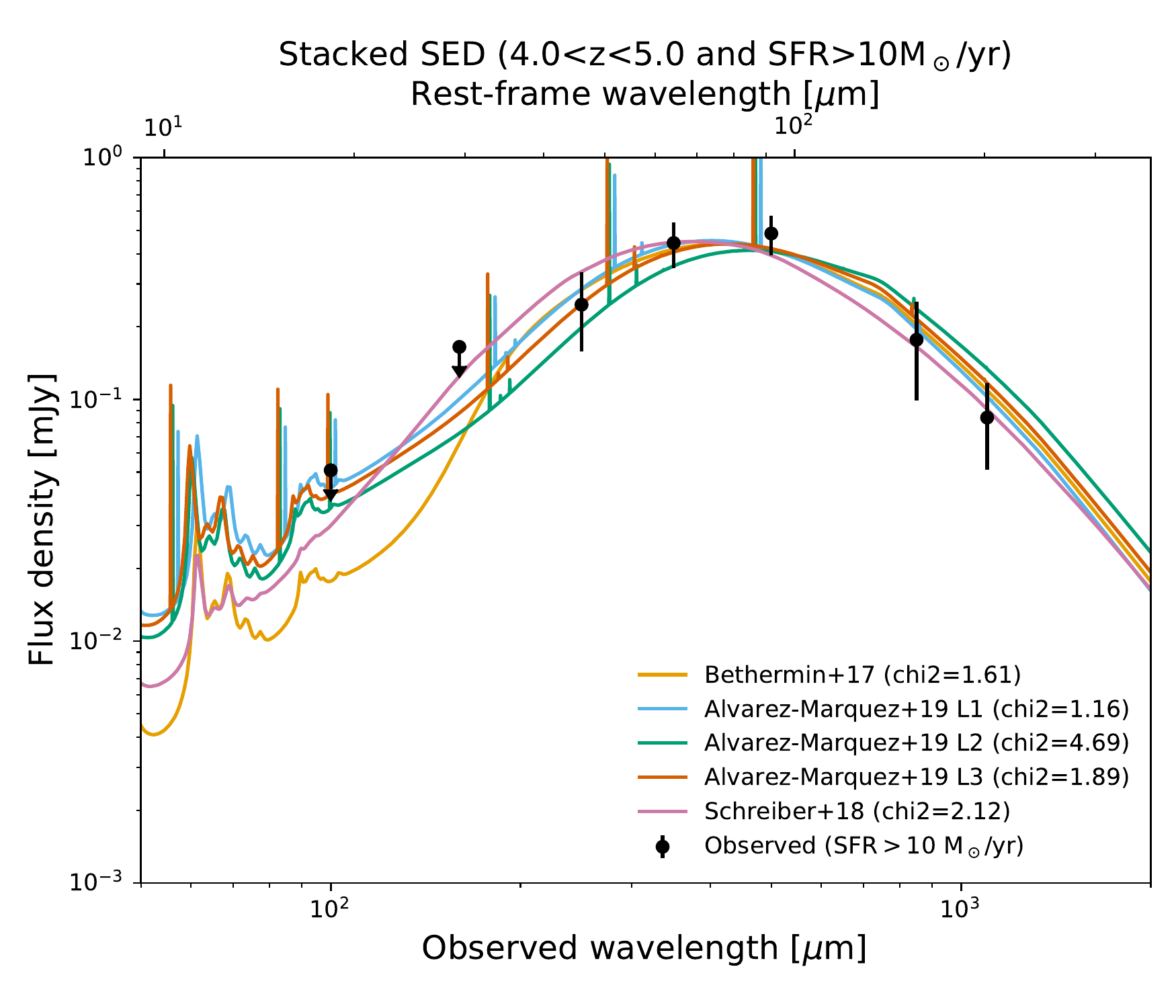}
\caption{Comparison of the chosen dust SED templates with the stacks of Herschel data at 4<z<5. The black symbols are stacked Herschel data and the solid lines are templates from the literature.}
\label{fig_templates_z4}
\end{figure*}

\begin{figure*}[h]
\ContinuedFloat
\captionsetup{list=off,format=cont}
\centering
\includegraphics[width=14cm]{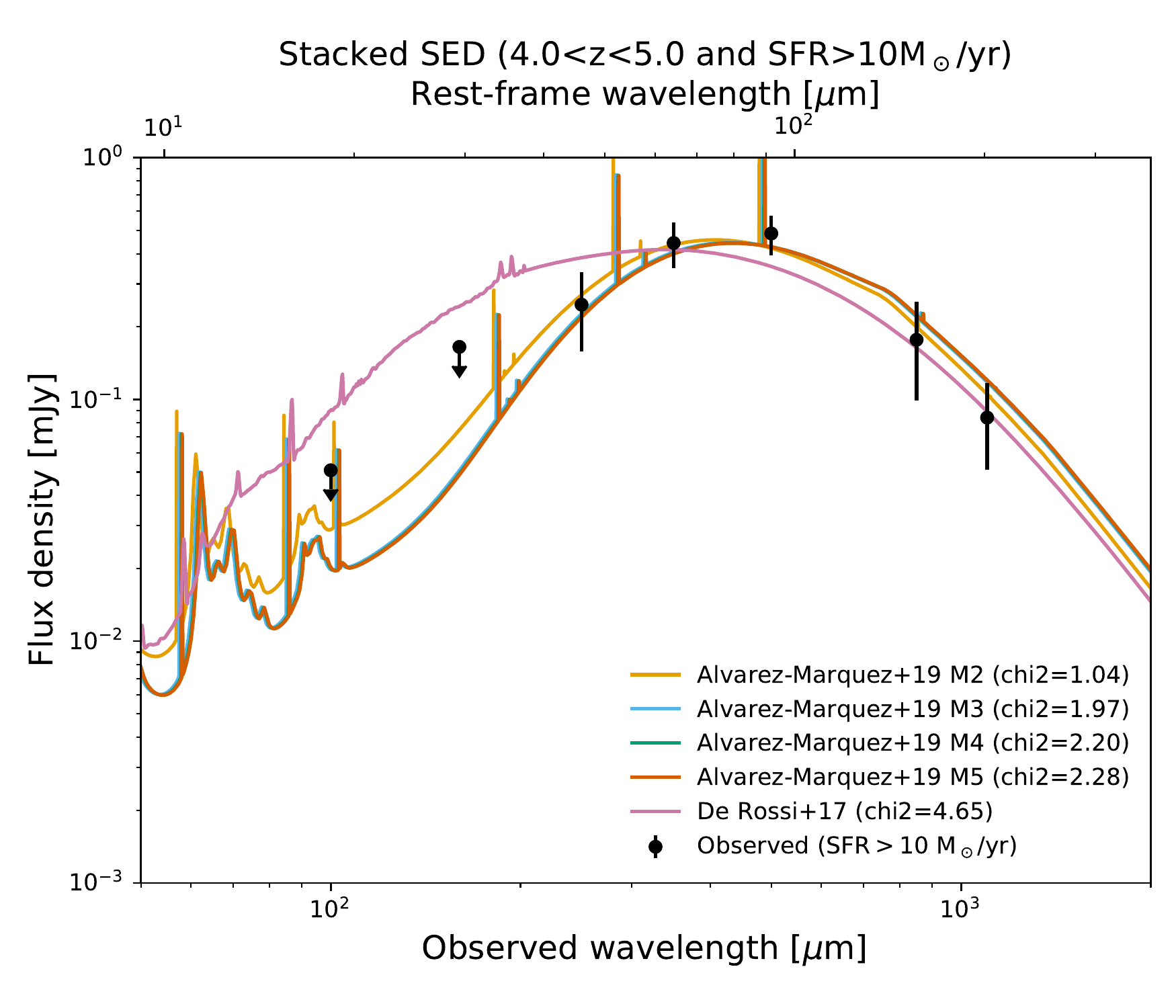}
\caption{}
\end{figure*}

\begin{figure*}[h]
\centering
\includegraphics[width=13cm]{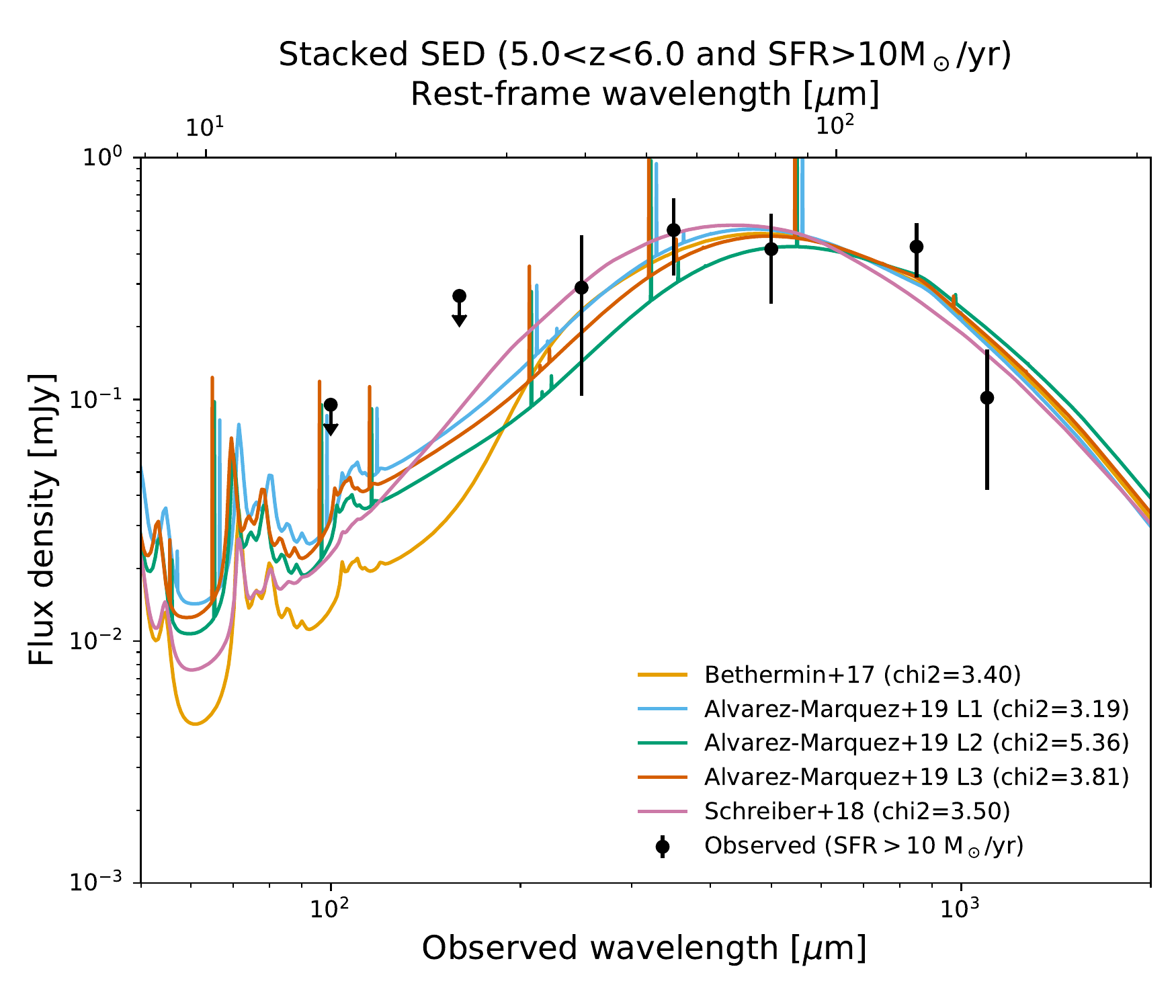}
\caption{Comparison of the chosen dust SED templates with the stacks of Herschel data at 5<z<6. The black symbols are stacked Herschel data and the solid lines are templates from the literature.}
\label{fig_templates_z5}
\end{figure*}

\begin{figure*}[h]
\ContinuedFloat
\captionsetup{list=off,format=cont}
\centering
\includegraphics[width=13cm]{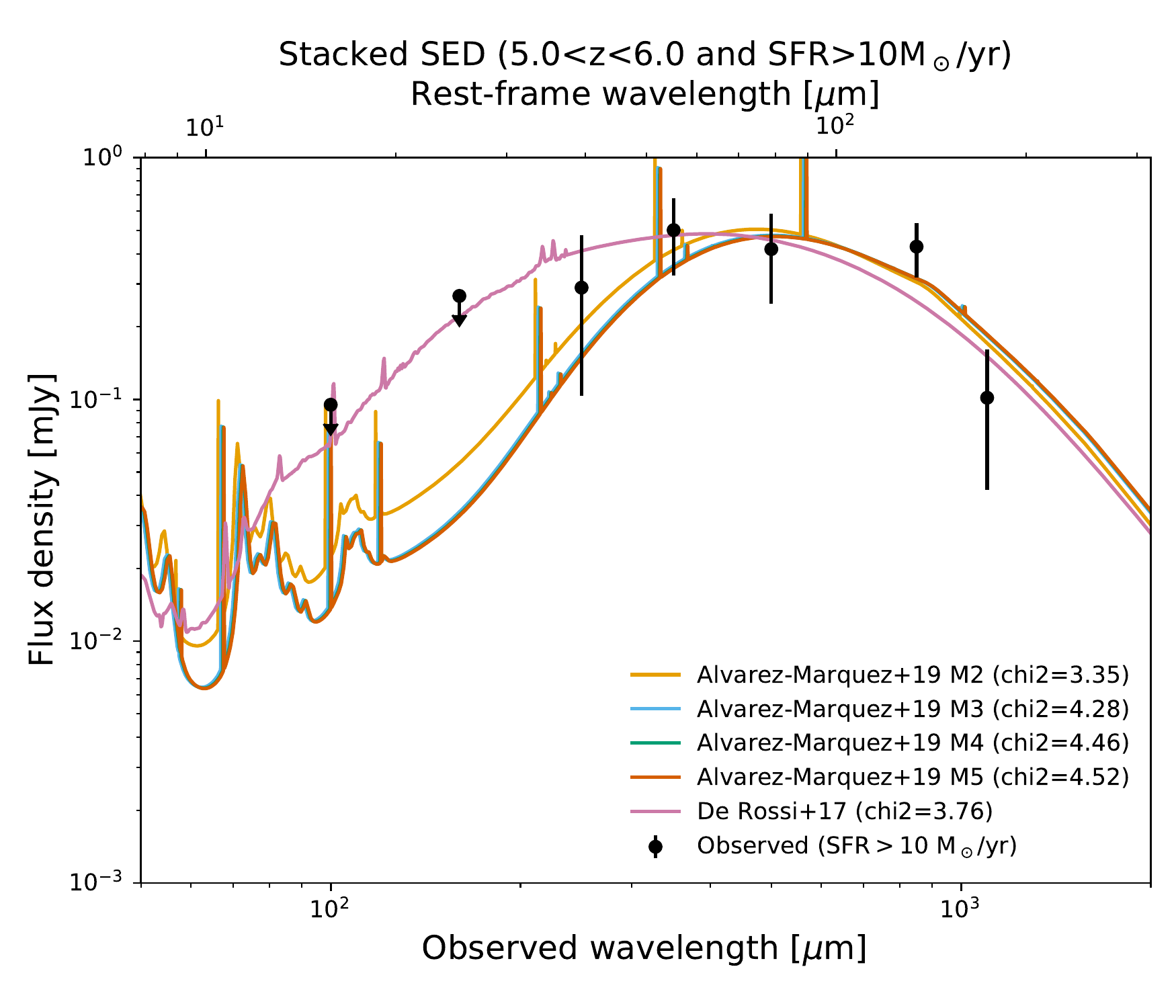}
\caption{}
\end{figure*}

\begin{figure*}[h]
\centering
\includegraphics[width=13cm]{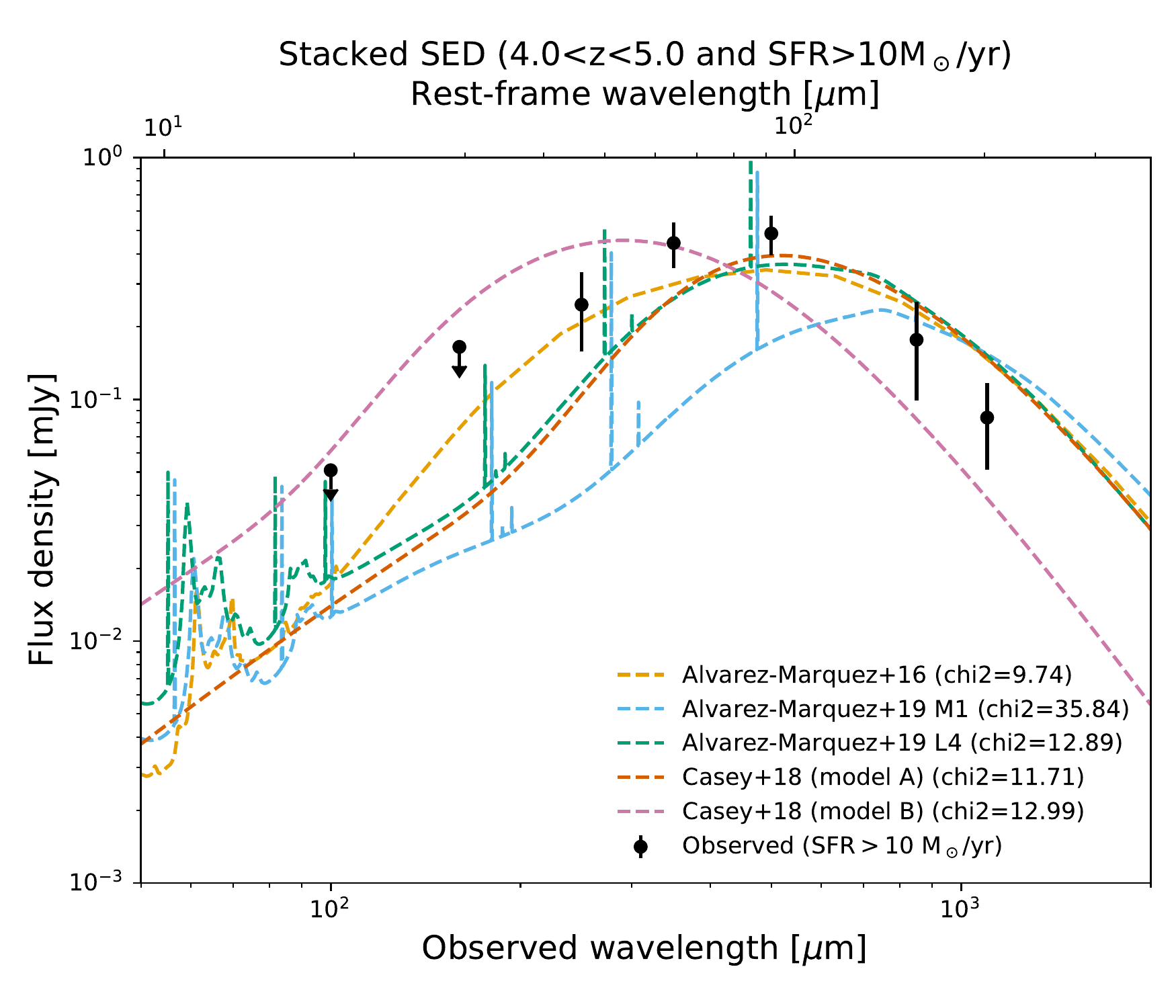}
\caption{Comparison of the discarded dust SED templates with the stacks of Herschel data at. The black symbols are stacked Herschel data and the dashed lines are templates from the literature.}
\label{fig_templates_discarded}
\end{figure*}

\begin{figure*}[h]
\ContinuedFloat
\captionsetup{list=off,format=cont}
\centering
\includegraphics[width=13cm]{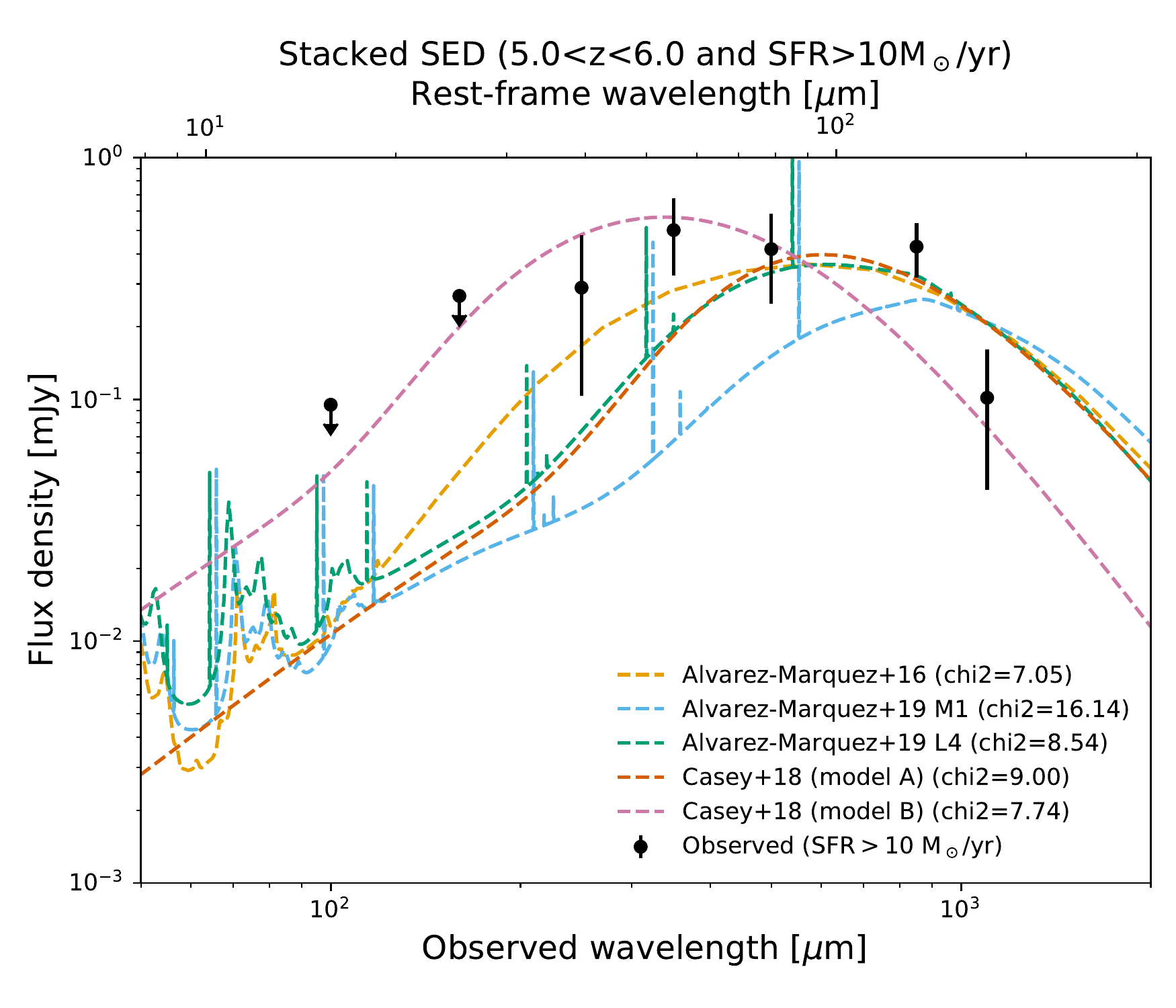}
\caption{}
\end{figure*}
 
\end{appendix}

\end{document}